\documentclass[aps,prb,twocolumn,groupedaddress,showpacs]{revtex4}

\usepackage[dvips]{graphicx}
\usepackage{amsmath}
\usepackage{amstext}
\usepackage{graphics}
\usepackage{exscale}
\usepackage{epsfig}
\usepackage[latin1]{inputenc}
\usepackage{bm}
\usepackage{bbm}

\usepackage{latexsym}

\renewcommand{\epsilon}{\varepsilon}

\newcommand{\integral}[3]{\!\int\limits_{#2}^{#3}\!\!{\rm d}#1\;}

\newcommand{\expval}[2]{ \langle  #1 #2\ \!\! \rangle}

\newcommand{\elcre}[2]{ c^{\dagger}_{#1,#2}}
\newcommand{\elann}[2]{ c_{#1,#2}}

\newcommand{\thGf}{{\cal G}}

\newcommand{\Imag}{\mathrm{Im}}
\newcommand{\Real}{\mathrm{Re}}

\newcommand{\hc}{\mathrm{h.c.}}

\begin{document}

\title{Field dependent quasiparticles in models of strongly correlated electrons}
\author{J. Bauer and A.C. Hewson}
\affiliation{Department of Mathematics, Imperial College, London SW7 2AZ,
  United Kingdom}
\date{\today} 
\begin{abstract}
In earlier work we showed how the low energy behavior
of the symmetric Anderson model in a magnetic field $H$ could be described in terms of  field dependent
renormalized quasiparticles. Here we extend the approach to the non-symmetric Anderson impurity model 
and to the infinite dimensional Hubbard model within the DMFT approach. We present NRG results
for the local spectral densities and the local longitudinal and transverse
dynamic spin susceptibilities for different parameter regimes and  a
sequence of values of the magnetic field. We calculate renormalized
parameters which characterize the quasiparticle excitations. Away from half-filling we find
quasiparticle weights, $z_\sigma(H)$, which  differ according to the spin type
$\sigma=\uparrow$ or  $\sigma=\downarrow$. Using the renormalized perturbation
theory, we show that the low energy features in  the NRG results can be well
described in the metallic phase  in terms of these field dependent
parameters. We also check Luttinger's theorem for the Hubbard model and find
it to be satisfied  in all parameter regimes  and for all values of the
magnetic field.  

 \end{abstract}

\maketitle

\section{Introduction}\par

A feature of strongly correlated electron systems, such as heavy fermions,
is their sensitivity  to an applied magnetic field, 
which makes  a magnetic field
 a useful experimental probe of strong correlation behavior.
 A manifestation of this sensitivity is the very 
large paramagnetic susceptibility observed in these systems. In terms of
Fermi liquid theory, the large paramagnetic susceptibility can be
interpreted as due to quasiparticles with exceptionally large effective
masses. These large effective masses arise from the scattering of
the electrons with the enhanced spin
fluctuations induced by the strong local Coulomb interactions.
An applied magnetic field suppresses the spin fluctuations causing a reduction
in the effective masses, which can be
seen experimentally in de Haas-van Alphen measurements \cite{JRCTDF87,GHTYF99}. Not only do
the effective masses depend on the magnetic field, they may also  differ for
the spin up and spin down electrons \cite{AUAO93,KSWA95}. Another feature that reflects
the enhanced sensitivity to an applied field is metamagnetic behavior, where
the spin susceptibility $\chi(H)$ in a finite field $H$ increases with the
field strength such that $d\chi(H)/dH>0$, which has been observed in some heavy
fermion compounds \cite{MCCSRC00}.  
As a related feature it has also been predicted that strong magnetic fields
can induce localization in narrow conduction bands \cite{Vol84,LGK94,JC00},
and this has been observed experimentally in quasi-two dimensional organic
conductors \cite{KIMK04}. 

In an earlier paper \cite{HBK06}, we looked at local models of strong
correlation behavior in a magnetic field and  showed how the quasiparticles
can be described by field-dependent renormalized parameters. We based our
approach on the single impurity  Anderson model\cite{And61}, which can be
characterized by the three independent parameters, $\epsilon_d$, the impurity
level, $\Delta$, the broadening of this level due to the hybridization with
conduction electrons, and $U$, the interaction at the impurity site. In the
absence of a magnetic field, the low energy quasiparticle excitations can be
described by an effective version of the same model with three corresponding
renormalized parameters, $\tilde\epsilon_d$,  $\tilde\Delta$, and $\tilde U$
\cite{HOM04,Hew93b,Hew05}.  We have shown that, in the presence of a  magnetic
field $H$, these  parameters can be taken as field-dependent,
$\tilde\epsilon_{d,\sigma}(H)$,  $\tilde\Delta_\sigma(H)$, and $\tilde
U(H)$. For the particle-hole symmetric model, we   calculated the
$H$-dependence explicitly from  numerical renormalization group (NRG)
calculations of the low energy excitations \cite{Hew05,HBK06}. The process of
de-renormalization of the quasiparticles with increase of mangetic field can be followed in these results.
The parameters which describe the quasiparticles change slowly on increasing
the field from zero and 
revert to their uncorrelated mean field values in the extreme high field limit. 
\par 
The renormalized parameters are not just a convenient way of describing the
low energy behavior; they  completely specify the model. A
renormalized perturbation theory (RPT) can be set up in which the free
propagators correspond to fully dressed quasiparticles
\cite{Hew93,Hew01}. This formalism is particularly effective for describing
the Fermi liquid regime, as only  diagrams up to second order have to be taken
into account to obtain asymptotically exact results for the $T=0$
susceptibilities, and the leading $T^2$ term in the conductivity. This
perturbation expansion is not restricted to the low energy and low 
temperature regime, and can be used for calculations on all energy scales. We
have  shown that a very good description of the $T=0$ spin and charge dynamics
for the Anderson model in the Kondo regime can be obtained  by summing the RPT
diagrams for repeated quasiparticle scattering \cite{Hew06}. The results give
an accurate description of the spin and charge susceptibilities for arbitrary
magnetic field values $H$, and for frequencies $\omega$ extending over a range
significantly larger than the Kondo temperature $T_{\rm K}$. The
Korringa-Shiba relation \cite{Shi75} and the sum rules for the spectral
density are satisfied.  
 \par 
In this paper we further extend this renormalized parameter approach to
infinite dimensional lattice models, concentrating particularly on the Hubbard
model. However, before doing so, we generalize some of our earlier results for
the Anderson model to situations away particle-hole symmetry. We do this not
only for the sake of completeness but  also so we can make a comparison with
the results for the lattice model in this regime.  
\par

\section{The non-symmetric Anderson model in a magnetic field}
The Hamiltonian for the Anderson model \cite{And61} is
\begin{eqnarray}
&&H_{\rm AM}=\sum\sb {\sigma}\epsilon\sb {\mathrm{d},\sigma} d\sp {\dagger}\sb
{\sigma}  
d\sp {}\sb {\sigma}+Un\sb {\mathrm{d},\uparrow}n\sb {\mathrm{d},\downarrow} \label{ham}\\
&& +\sum\sb {{ k},\sigma}( V\sb { k,\sigma}d\sp {\dagger}\sb {\sigma}
c\sp {}\sb {{ k},\sigma}+ V\sb { k,\sigma}\sp *c\sp {\dagger}\sb {{
k},\sigma}d\sp {}\sb {\sigma})+\sum\sb {{
k},\sigma}\epsilon\sb {{ k},\sigma}c\sp {\dagger}\sb {{ k},\sigma}
c\sp {}\sb {{
k},\sigma}, \nonumber
\end{eqnarray}
where $\epsilon_{\mathrm{d},\sigma}=\epsilon_{\rm d}-\sigma g\mu_{\rm B} H/2$
is the energy of the localized  level at an impurity site   in a magnetic
field $H$, $U$ the interaction at this local site, and $V_{k,\sigma}$ the
hybridization matrix element to a band of conduction electrons of spin
$\sigma$ with energy $\epsilon_{k,\sigma}-\sigma g_c\mu_{\rm B} H/2$, where
$g_c$ is the g-factor for the conduction electrons. When $U=0$ the local level
broadens into a resonance, corresponding to a localized quasi-bound state,
whose width depends on the quantity $ \Delta_\sigma(\omega)=\pi\sum\sb {k}|
V\sb {k,\sigma}|\sp 2\delta(\omega -\epsilon\sb { k,\sigma})$. For the
impurity model, where we are interested in universal features, it is usual to
take   a wide conduction band with a flat density of states so that
$\Delta_\sigma(\omega)$ becomes independent of $\omega$,
and can be taken as a constant $\Delta_\sigma$. In this wide band
limit $\Delta_\sigma(\omega)$ will be independent of the magnetic field
on the conduction electrons, so we can effectively put $g_c=0$.
When this is the case $\Delta_\sigma$ is usually taken to be a constant 
$\Delta$ independent of $\sigma$. 

\par
In the renormalized perturbation theory approach\cite{Hew93,Hew01} we cast the corresponding
Lagrangian for this model ${\cal L}_{\rm
AM}(\epsilon_{\mathrm{d},\sigma},\Delta,U)$  into  the form,
\begin{equation}
{\cal L}_{\rm AM}(\epsilon_{\mathrm{d},\sigma},
\Delta,U)={ \cal L}_{\rm AM}(\tilde\epsilon_{\mathrm{d},\sigma},
\tilde\Delta_\sigma,\tilde U)+ {\cal L}_{\rm
  ct}(\lambda_1,\lambda_2,\lambda_3),\label{lag}
\end{equation}
where the renormalized parameters,   $\tilde\epsilon_{\mathrm{d},\sigma}$ and 
$\tilde\Delta_{\sigma}$, are defined in terms of the self-energy
$\Sigma_{\sigma}(\omega)$ of the one-electron Green function for the impurity state,
\begin{equation}
G_{\sigma}(\omega)={1\over
    \omega-\epsilon_{\mathrm{d}\sigma}+i\Delta-\Sigma_\sigma(\omega)},
\label{gf}
\end{equation}
and are given by 
\begin{equation}
\tilde\epsilon_{\mathrm{d},{\sigma}}=z_{\sigma}(\epsilon_{\mathrm{d},{\sigma}} 
+\Sigma_\sigma(0)),\quad\tilde\Delta_{\sigma} =z_\sigma\Delta,
\label{ren1}
\end{equation} 
where $z_{\sigma}$ is given by
$z_{\sigma}={1/{(1-\Sigma_{\sigma}'(0))}}$.
The renormalized or quasiparticle interaction  $\tilde U$, is defined in terms
of the local total 4-vertex
$\Gamma_{\uparrow\downarrow}(\omega_1,\omega_2,\omega_3,\omega_4)$ at zero frequency,
  \begin{equation} 
\tilde U=z_{\uparrow}z_{\downarrow}\Gamma_{\uparrow\downarrow}(0,0,0,0).
\label{ren2}\end{equation}
It will be convenient to rewrite the spin dependent quasiparticle energies
in the form,
$\tilde\epsilon_{{\rm d},\sigma}=\tilde\epsilon_{\rm d}(h)-\sigma h\tilde\eta(h)$,
where
\begin{equation}
\tilde\epsilon_{\mathrm{d}}(h)={1\over
  2}\sum_\sigma\tilde\epsilon_{\mathrm{d},{\sigma}},\quad 
\tilde\eta(h)={1\over
  2h}\sum_\sigma \sigma\tilde\epsilon_{\mathrm{d},{\sigma}}
\end{equation}
where $\tilde\epsilon_{\rm d}(h)$ and $\tilde\eta(h)$ are both  even functions of the magnetic field
$h=g\mu_{\rm B}H/2$.\par
The renormalized perturbation expansion is in powers of the renormalized
interaction $\tilde U$ for the complete Lagrangian defined in equation
(\ref{lag}).  The counter term part of the Lagrangian $ {\cal L}_{\rm
  ct}(\lambda_1,\lambda_2,\lambda_3)$ essentially takes care of any 
overcounting.
\begin{figure}
\begin{center}
\includegraphics[width=0.45\textwidth,height=6cm]{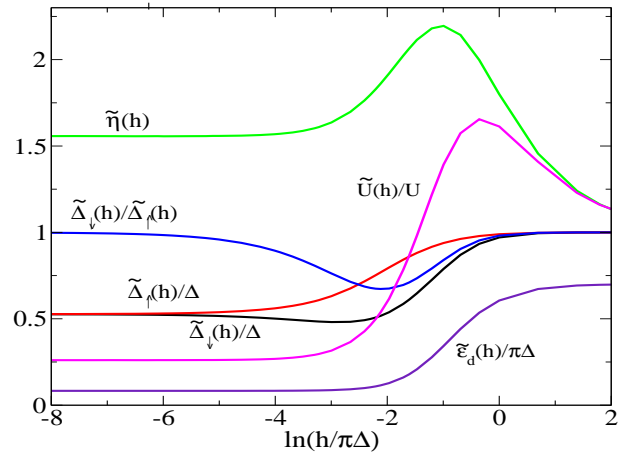}
\vspace*{-0.5cm}
\label{figure1}
\end{center} 
\caption{Plots of the renormalized parameters,
  $\tilde\Delta_{\uparrow}(h)/\Delta$, $\tilde\Delta_{\downarrow}(h)/\Delta$,
 $\tilde\epsilon_{\rm d}(h)/\pi\Delta$, $\tilde U(h)/U$, $\tilde\eta(h)$, for
 the asymmetric Anderson model, with $\pi\Delta=0.1$, $U/\pi\Delta=2$ and
 $\epsilon_d/\pi\Delta=-0.3$, as a function of the logarithm of the magnetic
 field $h/\pi\Delta$. The ratio
 $\tilde\Delta_{\downarrow}(h)/\tilde\Delta_{\uparrow}(h)$ is also shown.}
\end{figure}
The parameters, $\tilde\epsilon_{\mathrm{d},{\sigma}}$, $\tilde\Delta_\sigma$
 and $\tilde U$, have been taken to be the fully renormalized ones, and the
 counter term parameters, $\lambda_1$, $\lambda_2$ and $\lambda_3$, are
 required to cancel any further renormalization. They  are completely
determined by this condition.\par 
We can calculate the renormalized parameters by identifying the Lagrangian
${ \cal L}_{\rm AM}(\tilde\epsilon_{\mathrm{d},\sigma},
\tilde\Delta_\sigma,\tilde U)$ on the right hand side of equation (\ref{lag})
 with the leading
corrections to the fixed point in a numerical renormalization group calculation
for the same model \cite{Wil75,KWW80a,Hew93,Hew05}. Results for the general
model in the absence of a magnetic field \cite{HOM04}, and for the symmetric model in the
presence of a field, were given earlier \cite{HBK06}. There are some
significant differences for  the case of a magnetic field in situations
without  particle-hole symmetry which  
will describe here briefly. The main difference is that the wavefunction
renormalization factor  $z_\sigma(h)$  depends on the spin index $\sigma$,
and as a consequence so does the effective resonance width
$\tilde\Delta_{\sigma}(h)$, so the equations given earlier for the particle-hole symmetric
model \cite{HBK06} have to be generalized. The induced magnetization $M(h)$ is given by 
 $M(h)=g\mu_{\rm B}m(h)$, where
\begin{equation}
m(h)={1\over 2}(n_{{\rm d}\uparrow}-n_{{\rm d}\downarrow})={- 1\over 2\pi}\sum_\sigma
\sigma\,{\rm tan}^{-1}
\left(\frac{\tilde\epsilon_{{\rm d}\sigma}(h)}{\tilde\Delta_{\sigma}(h)}\right),
\label{magqp}
\end{equation}
which can be derived from the Friedel sum rule. The 
longitudinal  susceptibility  $\chi_l(h)$ (in units of $(g\mu_{\rm B})^2$)
is given by
\begin{equation}
\chi_l(h)=0.25(\tilde\rho_\uparrow(0,h)+\tilde\rho_\downarrow(0,h)+\tilde
U(h)\tilde\rho_\uparrow(0,h)\tilde\rho_\downarrow(0,h)),
\label{chil}
\end{equation}
where $\tilde\rho_\sigma(\omega,h)$ is the free quasiparticle density of states
given by 
\begin{equation}
\tilde\rho_\sigma(\omega,h)={1\over \pi}{\tilde\Delta_\sigma(h)\over
(\omega-\tilde\epsilon_{{\rm
    d},\sigma}(h))^2+\tilde\Delta^2_\sigma(h)}.
\label{rhosig}
\end{equation}
The corresponding transverse susceptibility $\chi_t(h)$ (zero applied field
limit in the transverse direction) is given by 
\begin{equation}
\chi_t(h)=\frac{m(h)}{2h}.
\label{chit}
\end{equation}
The total occupation of the impurity site $n(h)=(n_{{\rm d}\uparrow}+n_{{\rm
  d}\downarrow})$ can be derived similarly, and is given by
\begin{equation}
n(h)=1-{1\over 2\pi}\sum_\sigma {\rm tan}^{-1} 
\left(\frac{\tilde\epsilon_{{\rm
        d}{\sigma}}(h)}{\tilde\Delta_{\sigma}(h)}\right),
\label{nqp}
\end{equation}
and the local charge susceptibility  $\chi_c(h)$ is given by
\begin{equation}
\chi_c(h)=0.25[\tilde\rho_\uparrow(0,h)+\tilde\rho_\downarrow(0,h)-\tilde
U(h)\tilde\rho_\uparrow(0,h)\tilde\rho_\downarrow(0,h)] .
\label{chic}
\end{equation}
In figure 1 we display the renormalized parameters as a function of the magnetic
field on a log scale for the bare parameters $\epsilon_{\rm d}/\pi\Delta=-0.3$
and $U/\pi\Delta=2$, corresponding to a impurity occupation in the absence of
a field, $\langle
n_{d,\sigma}\rangle=n(0)/2\sim 0.35$. The overall trend is very similar to
that for the particle-hole symmetric case in the strong coupling regime. We do
see, however, that $\Delta_{\uparrow}(h)\ne\Delta_{\downarrow}(h)$, except
asymptotically as $h\to 0$ and $h\to\infty$. We note that, though
$\tilde\Delta_{\uparrow}(h)$ increases monotonically with increase of $h$,  
$\tilde\Delta_{\downarrow}(h)$ initially decreases. In this case, where the
impurity level is less than half-filled, the ratio
$\tilde\Delta_{\uparrow}(h)/\tilde\Delta_{\downarrow}(h)\ge 1$. This ratio is
reversed, so $\tilde\Delta_{\uparrow}(h)/\tilde\Delta_{\downarrow}(h)\le 1$,
when the impurity level is more than half-filled.  Shown in figure 2 is the
result for  $m(h)$ derived by substituting these parameters into equation
(\ref{magqp}), compared with results obtained by the direct evaluation of the
d-site occupation values in the ground state as determined from the NRG. 
\begin{figure}
\begin{center}
\includegraphics[width=0.45\textwidth,height=6cm]{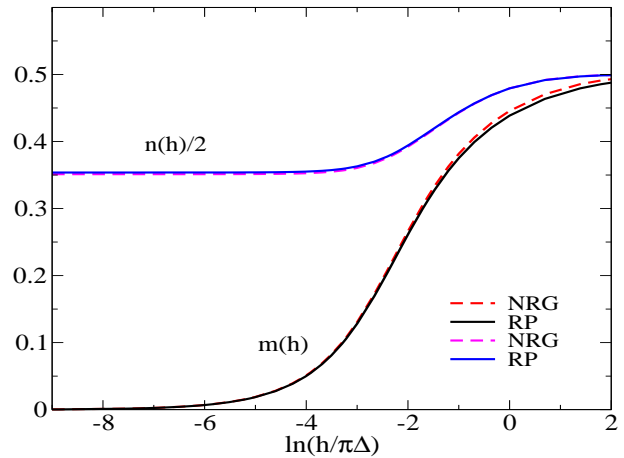}
\vspace*{-0.5cm}
\label{figure2}
\end{center} 
\caption{The induced magnetization $ m(h)$ as a function of the logarithm of
  the  magnetic field $h$ for the asymmetric Anderson model with the same set
  of parameters as given in figure 1. The dashed curve is that calculated from
  the direct evaluation of the occupation values from the NRG ground state,
  and the full curve is that deduced from the renormalized parameters in
  equation (\ref{magqp}). Also shown is the average occupation $n(h)/2$ as
  calculated from the NRG ground state (dashed curve) and the quasiparticle
  occupation values (full curve) as given in equation (\ref{nqp}).}
\end{figure}
There is a small but systematic difference, of the order of 2\%,  between the
two sets of results. The corresponding estimates of the  average occupation
number $n(h)/2$  as a function of magnetic field $h$ are shown in the same
figure. For this quantity the two sets of results are almost indistinguishable.
In the extreme large field limit the average occupation of the impurity
level tends to unity, as the majority spin level gets pulled further and
further below the Fermi level, and the impurity becomes completely polarized.
In this regime the average renormalized level, $\tilde\epsilon_{\rm d}$,
approaches the mean field value, 
$\tilde\epsilon_{\rm d}=\epsilon_{\rm d}+0.5Un(h)$, rather than the bare
value $\epsilon_{\rm d}$, while the other renormalized quantities approach
their bare values.\par
\begin{figure}
\begin{center}
\includegraphics[width=0.45\textwidth,height=6cm]{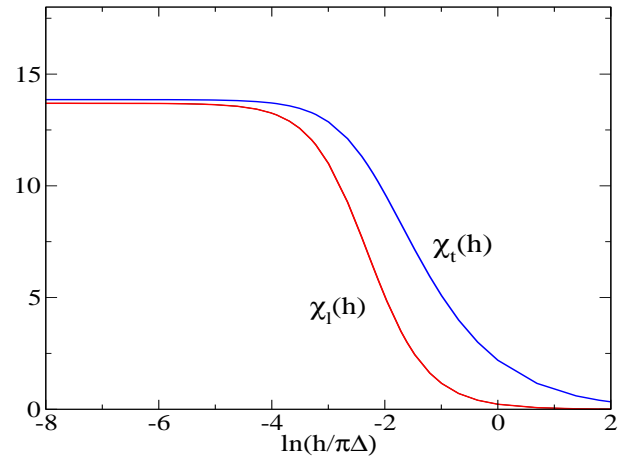}
\vspace*{-0.5cm}
\label{figure3}
\end{center} 
\caption{The longitudinal and transverse impurity site susceptibilities,
  $\chi_l(h)$ and  $\chi_t(h)$,  as a function of the logarithm of  the
 magnetic field $h$  for the asymmetric Anderson model with the same set of
 parameters as given in figure 1.  $\chi_l(h)$ is calculated from equation
 (\ref{chil}) and  $\chi_t(h)=m(h)/2h$. }
\vspace{0.5cm}
\label{fig:chitchil}
\end{figure}

The longitudinal and transverse spin susceptibilities,  $\chi_l(h)$ and
$\chi_t(h)$,  are plotted in figure \ref{fig:chitchil} as a function of the
 logarithm of  the  magnetic field $h$.   
$\chi_l(h)$ is calculated from equation (\ref{chil}) and  $\chi_t(h)=m(h)/2h$ 
as calculated from equation (\ref{magqp}). 
They should asymptotically converge to the same result in the limit $h\to 0$.
There seems to be a very small discrepancy, of about 1\%, between the two estimates in this
limit.\par

The generalization of our earlier results \cite{Hew06} for the dynamic longitudinal
susceptibility   $\chi_{l}(\omega,h)$, taking account of repeated quasiparticle-quasihole
scattering, is
\begin{eqnarray}
&&\chi_{l}(\omega,h)=  \label{lrrpt} \\
&&{\tilde\chi_{\uparrow\uparrow}(\omega,h)
 +\tilde\chi_{\downarrow\downarrow}(\omega,h)+4\tilde
 U_l(h)\tilde\chi_{\uparrow\uparrow}(\omega,h)\tilde\chi_{\downarrow\downarrow}(\omega,h)\over 2(
 1-4\tilde
 U^2_l(h)\tilde\chi_{\uparrow\uparrow}(\omega,h)\tilde\chi_{\downarrow\downarrow}(\omega,h))},
\nonumber
 \end{eqnarray}
where the analytic expression for $\tilde\chi_{\sigma\sigma}(\omega,h)$
is given in the Appendix. The zero frequency irreducible particle-hole vertex $\tilde U_l(h)$
in this scattering channel   
is given by
\begin{equation}
\tilde U_l(h)=
\frac{-1+\sqrt{[1+\tilde
    U^2(h)(\tilde\rho_\uparrow+\tilde\rho_\downarrow)]^2-\tilde
    U^2(h)(\tilde\rho_\uparrow-\tilde\rho_\downarrow)^2}}
  {2(\tilde\rho_\uparrow+\tilde\rho_\downarrow+\tilde U(h)
\tilde\rho_\uparrow\tilde\rho_\downarrow)},
\end{equation}
where we have simplified the notation,
$\tilde\rho_\sigma(0,h)=\tilde\rho_\sigma$. In the absence of a magnetic
field, or  with a magnetic field for the particle-hole symmetric model, 
    $\tilde\rho_\uparrow=\tilde\rho_\downarrow$, the result
simplifies to $\tilde U_l(h)=\tilde U(h)/[1+\tilde U(h)\tilde\rho(0,h)]$,
which is the value used in the earlier work\cite{Hew06}.\par

\begin{figure}
\begin{center}
\includegraphics[width=0.45\textwidth,height=6cm]{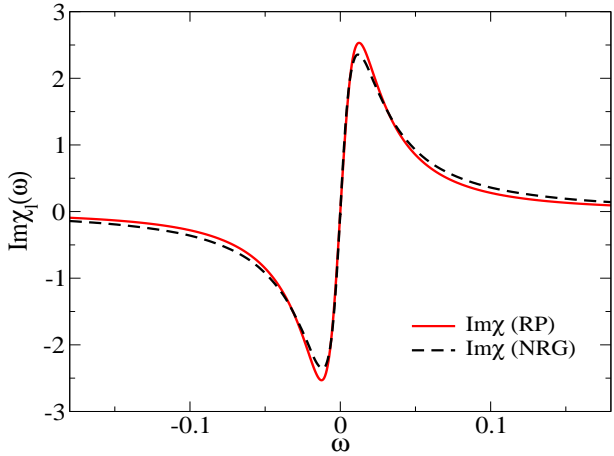}
\vspace*{-0.5cm}
\end{center} 
\caption{The imaginary part of the longitudinal dynamic susceptibility,
  $\chi_l(\omega,h)$  at $T=0$ as a function of $\omega$, for the asymmetric
  Anderson model for $h/\pi\Delta=0.001$, and the same set of parameters as
  given in figure 1. The dashed curve is calculated from a direct NRG
  calculation and the full curve from equation (\ref{lrrpt}) with the
  renormalized parameters. }
\label{imchilh0.001}
\end{figure}

\begin{figure}
\begin{center}
\includegraphics[width=0.45\textwidth,height=6cm]{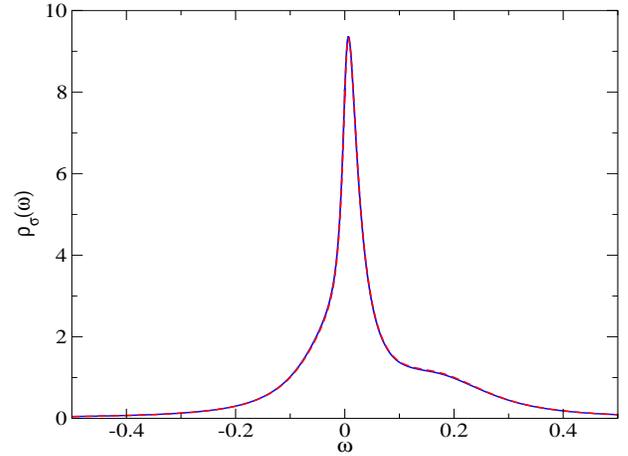}
\vspace*{-0.5cm}
\label{figure4}
\end{center} 
\caption{The spectral density for the spin up (majority) electrons
  $\rho_{\uparrow}(\omega,h)$  (full curve) and spin down (minority) electrons
  $\rho_{\downarrow}(\omega,h)$ (dashed curve) at $T=0$ as a function of
  $\omega$ for the same set parameters and magnetic field value as used for
  the plot of $\chi_l(\omega,h)$   in figure 4. The two curves are almost
  coincident on the scale shown.}
\label{rho0.0001}
\end{figure}

The imaginary part of the longitudinal dynamic susceptibility is shown in
figure \ref{imchilh0.001} for a magnetic field value $h/\pi\Delta=0.001$
for the same set of parameters, as for figures 2 and 3. 
The dashed curve is that from a direct NRG evaluation \cite{SSK89,CHZ94} and the full curve is that
calculated using equation (\ref{lrrpt}), with the corresponding renormalized parameters.
In the direct NRG evaluation we have used the improved method \cite{PPA06pre,WD06pre} of evaluating
the response functions with the complete Anders-Schiller basis \cite{AS05} so that the
sum rule for the total spectral density is satisfied exactly.
There is very good agreement between the two sets of results. The
Korringa-Shiba relation does not hold for the model without particle-hole
symmetry in the presence of a magnetic field. Where it  does hold, in the
absence of a field, or with particle-hole symmetry, the renormalized
perturbation expression satisfies it exactly.  The spectral densities,
$\rho_\uparrow(\omega)$  and  $\rho_\downarrow(\omega)$, as calculated from the NRG  for this field
value and the same set of parameters, are shown in figure \ref{rho0.0001}. 
For this value of the field there are  only a small differences between the two spectral
densities; they are almost coincident on the scale shown.

The dynamic transverse susceptibility $\chi_{t}(\omega,h)$, taking into
account the quasiparticle-quasihole repeated scattering, is given by 
\begin{equation}
\chi_{t}(\omega,h)={\tilde\chi_{\uparrow\downarrow}(\omega,h)
\over 1-\tilde
U_t(h)\tilde\chi_{\uparrow\downarrow}(\omega,h)},
\label{trrpt}
\end{equation}
where the analytic expression for $\tilde\chi_{\uparrow\downarrow}(\omega,h)$
is given in the Appendix. The zero frequency irreducible particle-hole vertex
$\tilde U_t(h)$ in this
channel is
\begin{equation}
\tilde U_t(h)={1\over m(h)}
\left\{{4\tilde h^2+(\tilde\Delta_\uparrow-\tilde\Delta_\downarrow)^2
\over 4\tilde h+{(\tilde\Delta_\uparrow-\tilde\Delta_\downarrow)
\over 2\pi m(h)}\,{\rm ln}\left({(\tilde
  h-\tilde\epsilon_d)^2+\tilde\Delta_\uparrow^2\over (\tilde
  h+\tilde\epsilon_d)^2+\tilde\Delta_\downarrow^2}\right)}-h\right\},
\end{equation}
where $\tilde h=h\tilde\eta(h)$, and  $m(h)$ is given in equation (\ref{magqp}).
This expression simplifies in the case,
$\tilde\Delta_\uparrow=\tilde\Delta_\downarrow$, to give $\tilde
U_t(h)=(\tilde h -h)/m(h)$, which is the same as that used
earlier\cite{Hew06}.\par
\begin{figure}
\begin{center}
\includegraphics[width=0.45\textwidth,height=6cm]{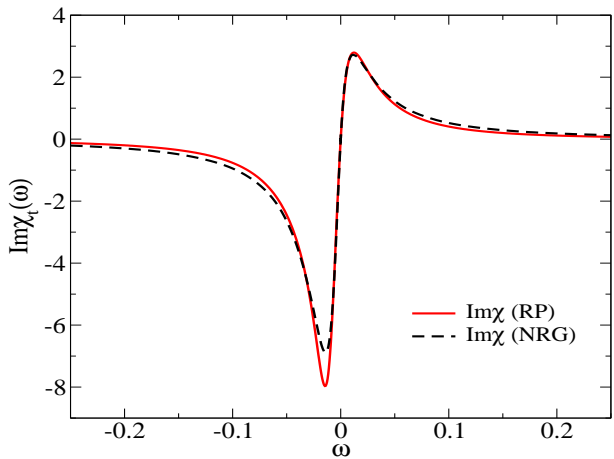}
\vspace*{-0.5cm}
\label{figure5}
\end{center} 
\caption{The imaginary part of the transverse dynamic susceptibility,  
$\chi_t(\omega,h)$  at $T=0$ as a function of $\omega$, 
for the asymmetric Anderson model for $h/\pi\Delta=0.04$ and
the same set of parameters as given in figure 1.
 }
\label{imchith0.04}
\end{figure}

 
Apart from a factor of 2, the dynamic transverse susceptibility in weak fields
is similar in form to the longitudinal one. In stronger fields, however,
one of the peaks in the imaginary part is suppressed while the other peak is
enhanced. Results are shown for the imaginary part of $\chi_{t}(\omega,h)$
in figure \ref{imchith0.04} for a field value $h/\pi\Delta=0.04$ and the same
parameters as for figures 2-5. 

\begin{figure}
\begin{center}
\includegraphics[width=0.45\textwidth,height=6cm]{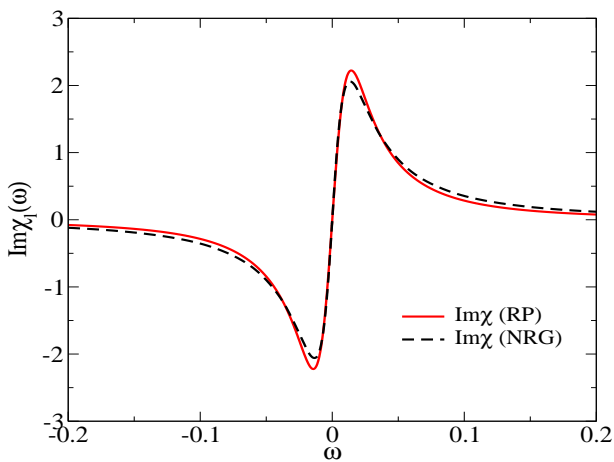}
\vspace*{-0.5cm}
\label{figure5}
\end{center} 
\caption{The imaginary part of the longitudinal dynamic susceptibility,
  $\chi_l(\omega,h)$  at $T=0$ as a function of $\omega$, for the asymmetric
  Anderson model for the same  parameters and magnetic field value
  ($h/\pi\Delta=0.04$) as given in figure 6.}
\label{imchilh0.04}
\end{figure}

\begin{figure}
\begin{center}
\includegraphics[width=0.45\textwidth,height=6cm]{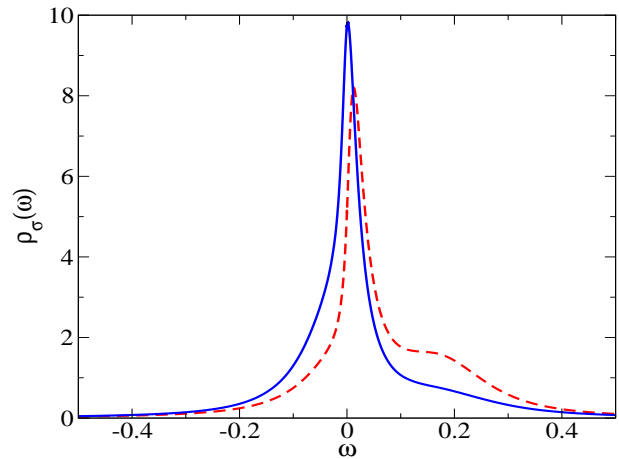}
\vspace*{-0.5cm}
\label{figure5}
\end{center} 
\caption{The spectral density for the spin up electrons
  $\rho_{\uparrow}(\omega,h)$  (full curve) and spin down
  $\rho_{\downarrow}(\omega,h)$ (dashed curve) at $T=0$ as a function of
  $\omega$ for the same set parameters and magnetic field value as used for
  the plot of $\chi_t(\omega,h)$ in figure 7.} 
\label{rhoh0.004}
\end{figure}

The full curve gives the results derived using
equation (\ref{trrpt}), and the dashed curve from a direct NRG calculation.
 The peak positions are in good agreement, and the slightly broader peak
from the NRG data can be attributed to the logarithmic broadening
used in the direct NRG evaluation. 
 There is a sum rule, that the total
spectral weight is equal to $-2m(h)$, which is satisfied precisely both in
the RPT result and also in the NRG calculation, as we have used the improved
prescription for the response functions based on the complete
Anders-Schiller basis.
The imaginary part of the corresponding dynamic longitudinal susceptibility
$\chi_{l}(\omega,h)$  for this value of the magnetic field is shown in figure
\ref{imchilh0.04}. 
Due to the stronger magnetic field, the peaks are somewhat suppressed as compared with the results
shown in figure  for $h/\pi\Delta=0.001$, but the overall features are very similar. 
The  spectral densities, $\rho_\uparrow(\omega)$ and
$\rho_\downarrow(\omega)$, for this field  ($h/\pi\Delta=0.04$)  are shown in
figure \ref{rhoh0.004}. 
As the polarization is much stronger for this higher magnetic field value,
there is now a marked difference between the two spectral densities for the
two spin types. \par
In summary, the results of this section for the impurity models away from
particle-hole symmetry display new features, such as the spin dependent
resonance widths in the quasiparticle density of states. We show that
the magnetisation and the static  response functions can be well described 
in terms of the field dependent renormalized parameters. We have used these parameters to
calculate the dynamic longitudinal and transverse spin susceptibilities in the
RPT formulae and we find excellent results when compared with those obtained from a
direct NRG calculation.

\section{Infinite dimensional Hubbard Model}
In this section we turn our attention to lattice models with a local
interaction terms and in particular to the infinite dimensional
Hubbard model. The Hubbard model has played a similar role for lattice models
as the Anderson model for impurity models, being the simplest model of its
type, where the interplay of kinetic energy and strong local interactions can
be studied.  

\subsection{Dynamical mean field approach and renormalized parameters}
The Hamiltonian for the model in a magnetic field is given by
\begin{equation}
H=-\sum_{i,j,\sigma}(t_{ij}\elcre {i}{\sigma}\elann
{j}{\sigma}+\hc)-\sum_{i\sigma}\sigma h n_{i\sigma}+U\sum_in_{i,\uparrow}n_{i,\downarrow}\label{hubm},
\end{equation}
where $t_{ij}$ are the hopping matrix elements between sites $i$ and $j$,
$U$ is the on-site interaction.\par
From Dyson's equation, the Green's function $G_{{\bf k},\sigma}(\omega)$ can be expressed in the form, 
\begin{equation}
G_{{\bf k},\sigma}(\omega) ={1\over
\omega+\mu_\sigma-\Sigma_{\sigma}({\bf k},\omega)-\epsilon({\bf k})},
\end{equation}
where $\Sigma_{\sigma}({\bf k},\omega)$ is the proper self-energy,
and $\epsilon({\bf k})=\sum_{\bf k}e^{-{\bf k}\cdot({\bf R}_i-{\bf R}_ j)}t_{ij}$,
 $\mu_{\sigma}=\mu+\sigma h$, with the chemical potential  $\mu$.
The simplification that occurs for the model in the infinite dimensional limit
is that $\Sigma_{\sigma}({\bf k},\omega)$  becomes
a function of $\omega$ only \cite{MV89,Mue89}. In this case the local Green's function  
$ G_{\sigma}^{\mathrm{loc}}(\omega)$ can be expressed in the form,
\begin{equation}
G_{\sigma}^{\mathrm{loc}}(\omega) =\sum_{\bf k}G_{{\bf k},\sigma}(\omega)=
\int d\epsilon\frac{D(\epsilon)}
{\omega+\mu_\sigma -\Sigma_{\sigma}(\omega)-\epsilon},
\label{gloc}
\end{equation}
where $D(\epsilon)$ is the density of states for the non-interacting model
($U=0$).  It is  possible convert this lattice problem into an effective
impurity one \cite{GKKR96} and  write this Green's function in the form, 
\begin{equation}
 G_{\sigma}^{\mathrm{loc}}(\omega) =
{1\over  \thGf_{0,\sigma}^{-1}(\omega) -\Sigma_{\sigma}(\omega)},
\label{locgf}
\end{equation}
where
\begin{equation}
  \thGf_{0,\sigma}^{-1}(\omega)=G_{\sigma}^{\mathrm{loc}}(\omega)^{-1}
  +\Sigma_{\sigma}(\omega).
\label{tgf}
\end{equation}
The Green's function  $ G_{\sigma}^{\mathrm{loc}}(\omega)$ can  be
identified with the Green's function
$ G_{\sigma}(\omega)$ of an effective Anderson model,  by
re-expressing $\thGf_{0,\sigma}^{-1}(\omega)$ as
\begin{equation}
\thGf_{0,\sigma}^{-1}(\omega)=\omega+\mu+\sigma h-K_{\sigma}(\omega),
\label{thgfK}
\end{equation}
so that
\begin{equation}
G_{\sigma}(\omega)={1\over
    \omega-\epsilon_{\mathrm{d}\sigma}-K_\sigma(\omega)-\Sigma_\sigma(\omega)},
\label{gfdmft}
\end{equation}
 with $\epsilon_{\mathrm{d}\sigma}=-\mu_\sigma$. The
function $K_\sigma(\omega)$ plays the role of a dynamical mean field
describing the effective medium surrounding the impurity. In the impurity case
in the wide band limit we have $K_{\sigma}(\omega)=-i\Delta$. Here as can be seen from
equations (\ref{tgf}) and (\ref{locgf}), $K_\sigma(\omega)$ is a
function of the self-energy $\Sigma_\sigma(\omega)$, and hence depends on
$\sigma$.  As this self-energy is identified with the
impurity self-energy, which  will depend on  the form taken for $K_\sigma(\omega)$, it is
clear that this quantity has to be calculated self-consistently. Starting from
an initial form for $K_\sigma(\omega)$, $\Sigma_\sigma(\omega)$ is
calculated using an appropriate 'impurity solver' from which 
$ G_{\sigma}^{\mathrm{loc}}(\omega)$ can be calculated using equation
(\ref{gloc}), and a new result for $K_\sigma(\omega)$ from equations
(\ref{tgf}) and (\ref{thgfK}). This $K_\sigma(\omega)$ serves as an input for
the effective impurity problem and the process is repeated until it converges 
to give a self-consistent solution. These equations constitute the dynamic mean field
theory (DMFT), and further details can be found in the review article of
Georges et al. \cite{GKKR96}.
\par
Of the many impurity solvers the most commonly used are the Monte Carlo, the
Exact diagonalization (ED) method and the NRG,  all of which have advantages
and disadvantages. Here, we wish to calculate the field dependent 
renormalized parameters to describe the quasiparticles, as we did for the
Anderson model, so we use the NRG approach. It is also the most accurate method
for calculations at $T=0$ and for the
low energy excitations. There has been a DMFT study of the static properties 
of a half-filled  Hubbard model in a magnetic
field  using the ED method by Laloux et al. \cite{LGK94}. 
The focus of our paper here, however, is rather different so there is little overlap
with this earlier work but, where there is, we make comment and compare with their results. 
  \par
We need to specify the density of states $D(\omega)$ of the non-interacting
infinite dimensional model, which is usually taken to be either for a
tight-binding hypercubic or Bethe lattice. Here we take the
 semi-elliptical  form corresponding to a Bethe lattice,
\begin{equation}
    D(\omega)=\frac{2}{\pi D^2}
\sqrt{D^2-(\omega+ \mu_0)^2}
\label{dos}
 \end{equation}
where $2D$ is the band width, with $D=2t$ for the Hubbard model, and $\mu_0$
the chemical potential of the free electrons. We choose
this form, rather than the Gaussian density of states of the hypercubic lattice,
as it has a finite bandwidth.\par 
In the NRG approach \cite{KWW80a} the conduction band is logarithmically discretized
and the model then converted into the form of a one dimensional tight
binding chain, coupled via an effective hybridization $V_\sigma$ to the impurity
at one end.  In this representation $K_\sigma(\omega)=|V_\sigma|^2
g_{0,\sigma}(\omega)$, where $g_{0,\sigma}(\omega)$ is the one-electron Green's
function for the first site of the isolated conduction electron chain.
If we use this form in equation (\ref{gfdmft}), and expand the self-energy
$\Sigma_\sigma(\omega)$ to first order in $\omega$, we can write the result in
the form,
\begin{equation}
G_{\sigma}(\omega)={z_\sigma\over
    \omega-\tilde\epsilon_{\mathrm{d}\sigma}-|\tilde
    V_\sigma|^2g_{0,\sigma}(\omega)+{\rm O}(\omega^2)},
\label{rgfdmft}
\end{equation}
where 
\begin{equation}
\tilde\epsilon_{\mathrm{d}\sigma}=z_\sigma(\epsilon_{\mathrm{d}\sigma}+\Sigma_\sigma(0)),\quad
|\tilde V_\sigma|^2={z_\sigma}| V_\sigma|^2, 
\label{rp}
\end{equation}
where $z_\sigma=1/(1-\Sigma'_\sigma(0))$, very similar to the result in
equation (\ref{ren1}).
We can interpret this result as a free quasiparticle  propagator, $\tilde
G_{0,\sigma}(\omega)$, given by
\begin{equation}
\tilde G_{0,\sigma}(\omega)={1\over
    \omega-\tilde\epsilon_{\mathrm{d}\sigma}-|\tilde V_\sigma|^2g_{0,\sigma}(\omega)},
\label{qpgfdmft}
\end{equation}
and ${z_\sigma}$ as the quasiparticle weight.
A similar Fermi liquid expansion in (\ref{gloc}) leads to 
\begin{equation}
\tilde G_{0,\sigma}^{\mathrm{loc}}(\omega) =\int d\epsilon\frac{D(\epsilon/z_{\sigma})}
{\omega+\tilde\mu_{0,\sigma} -\epsilon},
\label{gqploc}
\end{equation}
where $\tilde\mu_{0,\sigma}=z_{\sigma}(\mu_{\sigma}-\Sigma_{\sigma}(0))$.
In the DMFT approach we identify
\begin{equation}
  \tilde G_{0,\sigma}^{\mathrm{loc}}(\omega)=\tilde G_{0,\sigma}(\omega),
\end{equation}
which specifies the form of $g_{0,\sigma}(\omega)$ in (\ref{qpgfdmft}) and
and yields $\tilde\mu_{0,\sigma}=-\tilde\epsilon_{\mathrm{d}\sigma}$.
We can  define a density of states $\tilde \rho_{0,\sigma}(\omega)$ for the free
quasiparticles via $\tilde \rho_{0,\sigma}(\omega)=-{\rm Im}\tilde
G_{0,\sigma}(\omega+i\delta)/\pi$. For the Bethe lattice, this
takes the form of a band with renormalized parameters,
\begin{equation}
   \tilde \rho_{0,\sigma}(\omega)=\frac{2}{\pi\tilde D_{\sigma}^2}\sqrt{\tilde 
   D_{\sigma}^2-(\omega+\tilde \mu_{0,\sigma})^2}.
\label{rdos}
 \end{equation}
where $\tilde D_{\sigma}=z_{\sigma}D$. 

We can deduce the renormalized parameters $\tilde\epsilon_{\mathrm{d}\sigma}$
and $\tilde V_\sigma$ from the low lying excitations calculation from the 
the NRG using a generalization of the method given in an earlier paper
\cite{HOM04}. The quasiparticle weight $z_\sigma$ is then obtained from the
relation $z_\sigma=|\tilde V_\sigma/ V_\sigma|^2$ in equation (\ref{rp}).\par

We can define a quasiparticle occupation number
 $ \tilde n^0_{\sigma}$ by integrating this density of states up to the Fermi level,
 \begin{equation}
   \tilde n^0_{\sigma}=\integral{\omega}{-\infty}{0}\tilde
   \rho_{0,\sigma}(\omega).
\label{qpocc}
 \end{equation}
Using Luttinger's theorem \cite{Lut60} it is possible to relate this free quasiparticle
occupation number  $ \tilde n^0_{\sigma}$ to the expectation value of the 
occupation number $n_{\sigma}$  in the interacting system at $T=0$.
Using the quasiparticle density of states in equation (\ref{rdos}),
we can rewrite equation (\ref{qpocc}) as
 \begin{equation}
   \tilde n^0_{\sigma}=\integral{\omega}{-\infty}{\infty} D(\omega)
\theta(\mu_\sigma-\Sigma_\sigma(0)-\omega),\label{nocc}
 \end{equation}
where $\theta(\omega)$ is the Heaviside step function and $D(\omega)$ as given
in equation (\ref{dos}). Then from
Luttinger's result the right-hand side of equation (\ref{nocc})
is equal to $n_{\sigma}$. We then have the result,
 \begin{equation}
   \tilde n^0_{\sigma}= n_{\sigma},
 \end{equation}
that the  occupation for electrons of spin $\sigma$ is equal to
the number of  free quasiparticle of spin $\sigma$, as calculated from
equation (\ref{qpocc}). It should be noted that there is no simple
generalization of the $h=0$ result, $\mu-\mu_0=\Sigma(0)$, in the spin
polarized case. 
\par
We can also calculate the local longitudinal and transverse dynamic spin
susceptibilities, $\chi_l(\omega)$ and  $\chi_t(\omega)$, for this model using  equations, (\ref{lrrpt}) and
  (\ref{trrpt}),  derived earlier
in the impurity case.  The free quasiparticle susceptibilities
$\tilde\chi_{\sigma,\sigma'}(\omega)$ required are calculated using the local
free quasiparticle density of states $\tilde\rho_{0,\sigma}(\omega)$ given in
equation (\ref{rdos}). We can calculate the local on-site quasiparticle
interaction $\tilde U$ as in the impurity case, but
 we do not  have the simple formula relating $\tilde U$ to $\chi_l(0)$ and
$\chi_t(0)$ 
that enabled us to deduce the irreducible quasiparticle interactions
$\tilde U_l$ and $\tilde U_t$; the impurity formula we used earlier 
is only valid in the wide band limit. To determine  $\tilde U_l$ and $\tilde
U_t$ in the lattice case we use
 the condition that  ${\rm Re}\chi_l(\omega)$ and
${\rm Re}\chi_t(\omega)$  fit the NRG result at the single
point $\omega=0$. We can then compare the results based on these RPT
formulae, which take into account the repeated quasiparticle scattering,
with the NRG results
 over the whole frequency
range.\par
Having covered the basic theory, we are now in a position to survey
the results for the Hubbard model in different parameter regimes.

\subsection{Results at Half-filling}

We present the results at half-filling for three main parameters regimes where we find
qualitatively different behavior. The results in all cases will be for a
Bethe lattice with a band width  $W=2D=4$, setting $t=1$. In concentrating on
the field induced polarization, we do not include the possibility of
antiferromagnetic ordering. The regimes are a relatively weak coupling regime
(a) where $U$ is smaller than the band width, an intermediate coupling regime
(b)  with $W<U<U_c$, where $U_c$ is the value at which a Mott-Hubbard gap
develops in the absence of a magnetic field ($U_c\approx 5.88$), and (c) a
strong coupling regime  with $U>U_c$.\par

\begin{figure}[!htbp]
\centering
\includegraphics[width=0.45\textwidth]{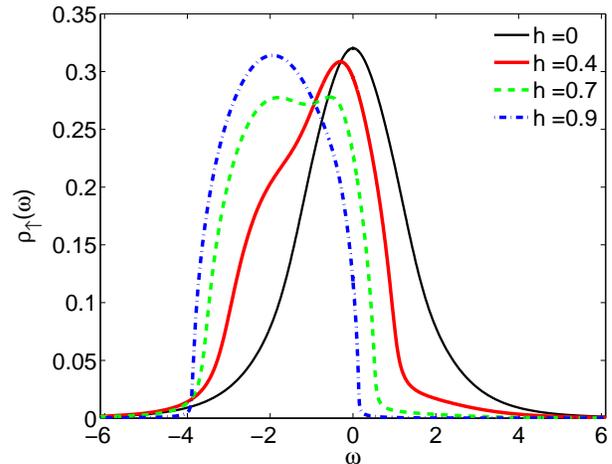}
\vspace*{-0.5cm}
\caption{The local spectral density for the majority spin $\rho_\uparrow(\omega)$
  for $U=2$ and various fields $h$. } 
\label{doswchdep}
\end{figure}

\begin{figure}[!htbp]
\centering
\includegraphics[width=0.45\textwidth]{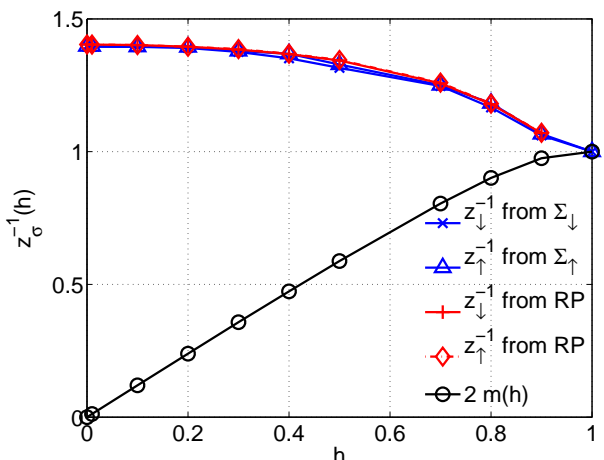}
\vspace*{-0.5cm}
\caption{The inverse of the quasiparticle weight $z_\sigma(h)$ calculated
  from renormalized parameters (RP) and directly from the self-energy and the
  magnetization $m(h)$ for $U=2$ and various fields $h$. }
\label{qpweightcompU2}
\end{figure}

The first plot in figure \ref{doswchdep} gives the spectral densities for the
majority spin electrons $\rho_\uparrow(\omega)$ for various magnetic field values
in the weakly correlated regime, $U=2$. 
We can see clearly that, for increasing magnetic field, more and more spectral
weight is shifted to lower energies  (the opposite happens for the other spin
component, which is not displayed here). Above $h\simeq 1.0$ the system
is completely polarized, $2m=1$. This extreme high field limit corresponds to
an insulator; there is a gap of the magnitude  $\Delta_g(h)=2h+U-W$  between the 
upper (minority) and lower (majority) bands, which both have the
semi-elliptical form  as for  non-interacting system  with $W=4$. The inverse
of the quasiparticle weight  $z_\sigma(h)$, which corresponds to the
enhancement of the effective mass $m^*_{\sigma}(h)=m/z_{\sigma}(h)$,
is shown as a function of $h$ in fig. \ref{qpweightcompU2}. 
The renormalized parameters (RP) $\tilde \epsilon_{\rm d,\sigma}$ and
 $|\tilde V_{\sigma}|^2$ were calculated from the NRG low energy excitations,
as described in the previous section, and the values of $z_\sigma(h)$ deduced
using equation (\ref{rp}). 
We also calculated $z_\sigma(h)$ from the numerical derivative of the NRG calculated self-energy 
$\Sigma_\sigma(\omega)$ for both spin up and spin down electrons. 
All values agree very well and $z_\uparrow(h)=z_\downarrow(h)$, as expected in the case
with particle-hole symmetry. The method based on the renormalized parameters
is only applicable in the metallic regime.  
The values of $z_\sigma(h)$  increase from about $0.75$ to $1$, which
corresponds to a progressive ``de-renormalization'' 
of the quasiparticles with increasing field, as observed earlier for the impurity model
\cite{HBK06}. Since the interaction term in the Hubbard model acts only for
opposite spins it is clear that there is no renormalization  when the system is
completely polarized with one band fully occupied and the
other empty. The expectation value of the double occupancy
$\expval{n_{\uparrow}n_{\downarrow}}{}$ decreases with increasing field, which
further illustrates why the interaction term becomes less important for larger fields.
 \par

\begin{figure}[!htbp]
\centering
\includegraphics[width=0.45\textwidth]{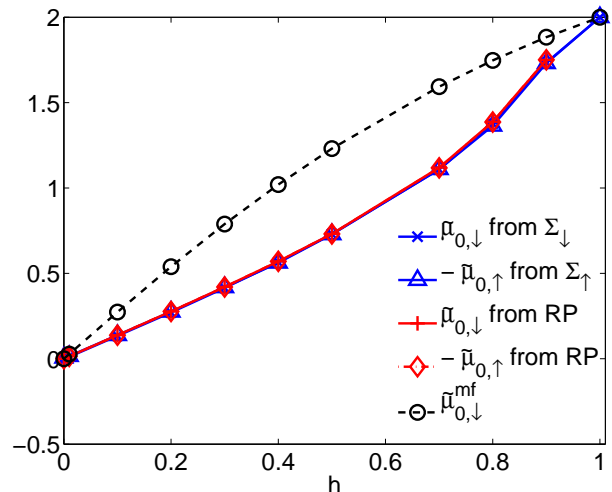}
\vspace*{-0.5cm}
\caption{The renormalized chemical potential $\tilde\mu_{0,\sigma}(h)$ calculated
  from renormalized parameters (RP) and directly from the self-energy for
  $U=2$ and various fields $h$. } 
\label{tmucompU2}
\end{figure}

We can also follow the field dependence of the renormalized chemical potential
$\tilde\mu_{0,\sigma}(h)$ as shown in figure \ref{tmucompU2}. It is shown
deduced from the renormalized parameter (RP) $\tilde \epsilon_{\rm d,\sigma}$
and as calculated directly from the self-energy. The agreement is very good
over the full range of magnetic fields. 

\begin{figure}[!htbp]
\centering
\includegraphics[width=0.45\textwidth]{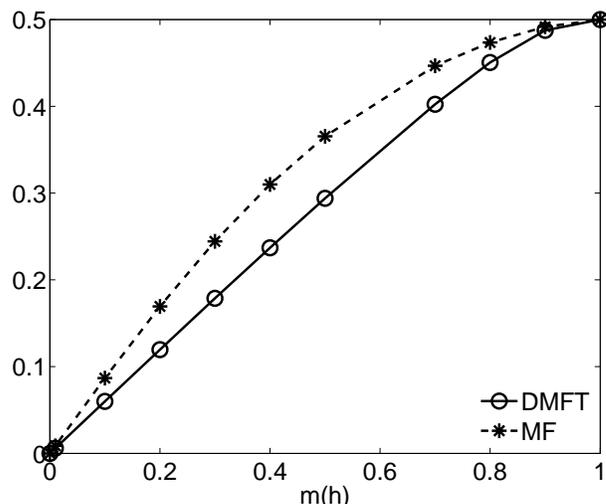}
\vspace*{-0.5cm}
\caption{The magnetization in the mean field approximation compared with the DMFT
  result for $U=2$ and for the full range of magnetic fields $h$. }
\label{magcompU2}
\end{figure}

Mean field theory is valid for very weak interactions, so we compare 
our results for  $\tilde\mu_{0,\sigma}(h)$ for $U=2$, with the
 mean field value  $\tilde\mu^{\rm mf}_{0,\sigma}=\mu+\sigma h
-Un^{\rm mf}_{-\sigma}$ in figure \ref{tmucompU2}.
The results coincide for $h=0$, when $\tilde\mu^{\rm
  mf}_{0,\sigma}=0$ and when the system becomes 
fully polarized at large field values, $\tilde\mu^{\rm
  mf}_{0,\sigma}=-\sigma(U+h)$, but in general  $\tilde\mu^{\rm
  mf}_{0,\sigma}>\tilde\mu_{0,\sigma}(h)$. We also compare the mean field (MF)
result for the field dependence of the magnetization $m(h)$  with the one
obtained in the DMFT calculation in figure \ref{magcompU2}. 
The general behavior is similar, but the mean field theory without quantum
fluctuations overestimates the magnetization, as one would expect.\par

\begin{figure}[!htbp]
\centering
\includegraphics[width=0.45\textwidth]{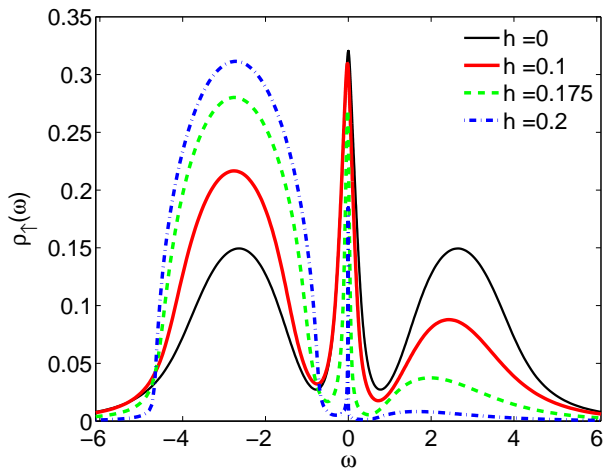}
\vspace*{-0.5cm}
\caption{The local spectral density for the majority spin $\rho_\uparrow(\omega)$
  for $U=5$ and various fields $h$. } 
\label{dosimchdep}
\end{figure}

\begin{figure}[!htbp]
\centering
\includegraphics[width=0.45\textwidth]{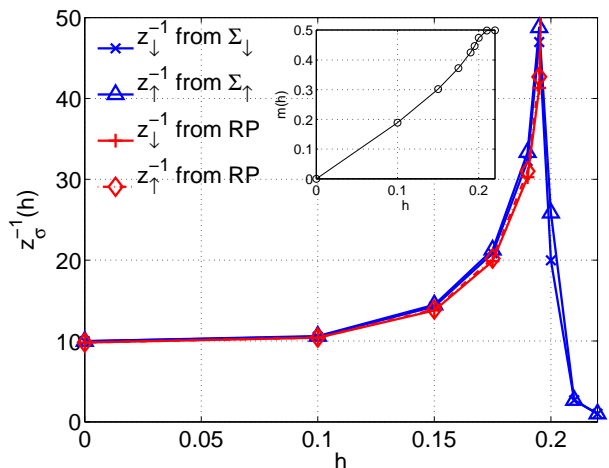}
\vspace*{-0.5cm}
\caption{The inverse of the quasiparticle weight $z_\sigma(h)$ calculated
  from renormalized parameters (RP) and directly from the self-energy for
  $U=5$ and various fields $h$. The inset shows the
  magnetization $m(h)$.   }
\label{qpweightcompU5}
\end{figure}

In the next plot in figure \ref{dosimchdep}, where $U=5$, we show   typical behavior in the
intermediate coupling  regime. 
Similar to the weak coupling regime, we find a shift of spectral
weight towards lower energy for the majority spin. There is, however, a
difference in the way this happens due to the initial three peak structure,
namely the quasiparticle peak in the middle gets narrower for increasing field
and finally vanishes in the polarized phase. The quasiparticle weight,
which is shown in figure \ref{qpweightcompU5},  reflects
this behavior by decreasing to zero  with increasing field signaling
heavy quasiparticles. When the material is polarized, however, the $z_\sigma(h)$ reverts
to 1, which corresponds to the band insulator as before. This approach to
the fully polarized localized state in high fields contrasts with that found 
in the weak coupling regime. It gives rise to metamagnetic behavior in this
parameter regime.

To  illustrate further this different response to a magnetic field, the real
part of the local longitudinal dynamic spin susceptibility $\chi_l(\omega,h)$ as a function 
of $\omega$ is shown for various values of $h$. It can be seen
that the local susceptibility $\chi^{\mathrm{loc}}(h)={\rm Re}\,\chi_l(0,h)$ in this regime increases
with $h$ so that $\partial\chi^{\mathrm{loc}}(h)/\partial h >0$. This can also be seen in the
curvature of the magnetization shown in the inset of figure \ref{qpweightcompU5}.
This is behavior characteristic of a  metamagnetic
transition and related to the magnetic field induced metal-insulator transition.\par

\begin{figure}[!htbp]
\centering
\includegraphics[width=0.45\textwidth]{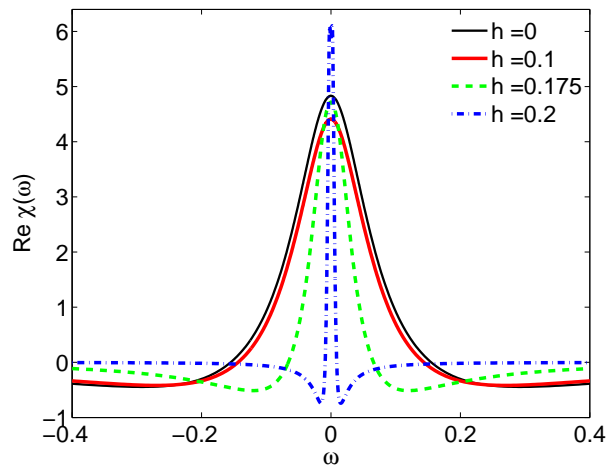}
\vspace*{-0.5cm}
\caption{The real part of the local longitudinal dynamic spin susceptibility for $U=5$ and
  various fields $h$.} 
\label{suscwchdep}
\end{figure}

\begin{figure}[!htbp]
\centering
\includegraphics[width=0.45\textwidth]{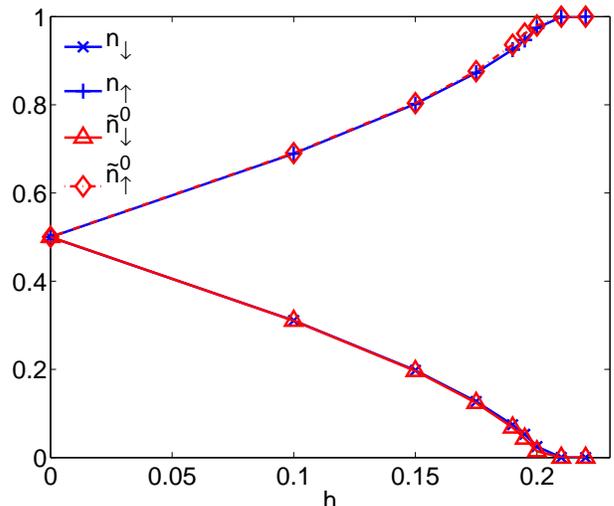}
\vspace*{-0.5cm}
\caption{The comparison of the spin dependent occupation numbers $\tilde
  n_\sigma^0$ and $n_\sigma$ corresponding to Luttinger's theorem in a magnetic
  field for $U=5$ and the range of fields $h$.  }
\label{luttU5}
\end{figure}

We can also check the Luttinger theorem in a magnetic field, as discussed
in the previous section, by comparing the  values of $\tilde n_\sigma^0$, 
deduced from integrating the quasiparticle density of states with the value
of  $n_\sigma$ calculated from the direct NRG evaluation in the ground state.
These results are shown in figure \ref{luttU5}. It can be seen that 
there is excellent agreement between the results of these two different
calculations,  $\tilde n_\sigma^0=n_\sigma$, so that
 Luttinger's theorem is satisfied for all values of the magnetic field in this
intermediate coupling regime. \par

\begin{figure}[!htbp]
\centering
\includegraphics[width=0.45\textwidth]{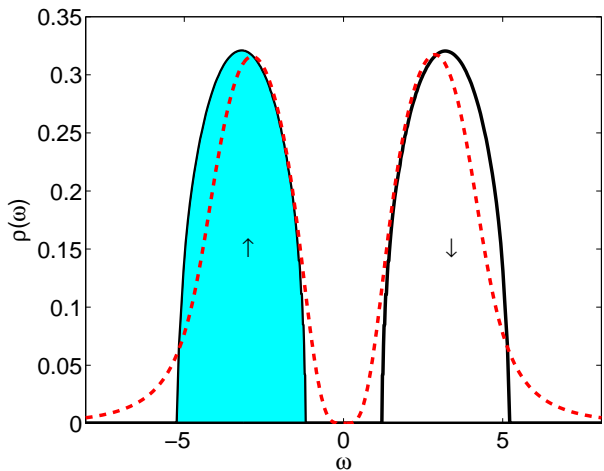}
\vspace*{-0.5cm}
\caption{The total local spectral density $\rho(\omega)$ for $U=6$ for $h=0$
  (dashed line), Mott insulator,  and   $h=0.2$ (full line), fully polarised
  band insulator.  }
\label{dosmi}
\end{figure}

Finally we consider the strong coupling regime with $U>U_c$,
where for $h=0$ the spectral density has a Mott-Hubbard gap
so that for half-filling the system is an insulator. 
The electrons will be localized with free magnetic moments coupled by
an effective antiferromagnetic exchange $J\sim t^2/U$. In fields
such that $h>J$, the system polarizes completely as can be seen in figure
\ref{dosmi} where we show the total density of states
$\rho(\omega)=\rho_{\uparrow}(\omega) +\rho_{\downarrow}(\omega)$ for $h=0$
and $h=0.2$. 
For smaller fields, such that $h<J$, we do not find a convergent
solution to  the DMFT equations, and the iterations oscillate between local
states which are either completely full or empty. We interpret this  as due to
the tendency to antiferromagnetic order which in a weak field, due to the
absence of anisotropy,  will be almost perpendicular to the applied field in
the x-y plane with a slight canting of the spins in the z-direction (spin
flopped phase). In this calculation no allowance has been made for this type of
ordering, but this state can be well described using an effective Heisenberg
model for the localized moments.

\subsection{Quarter Filled Case}

\begin{figure}[!htbp]
\centering
\includegraphics[width=0.45\textwidth]{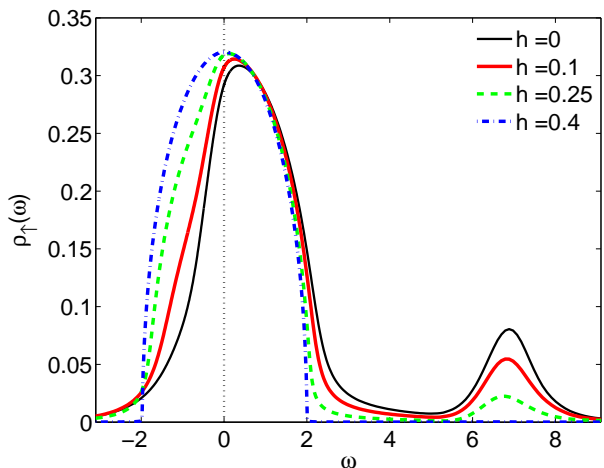}
\vspace*{-0.5cm}
\caption{The local spectral density for the majority spin $\rho_{\uparrow}(\omega)$
  for $U=5$, $x=0.5$ and various fields $h$. }
\label{dosdopU5u}
\end{figure}

\begin{figure}[!htbp]
\centering
\includegraphics[width=0.45\textwidth]{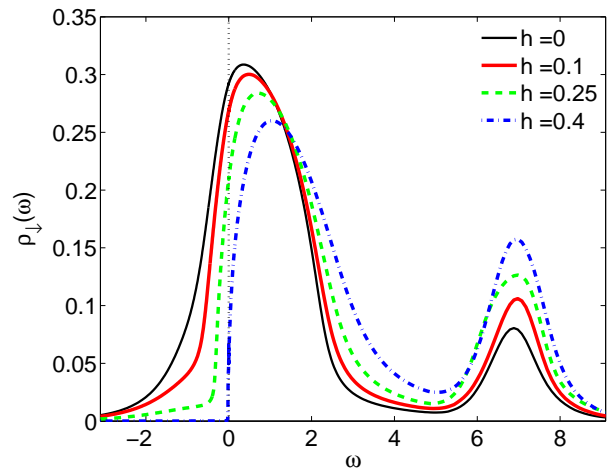}
\vspace*{-0.5cm}
\caption{The local spectral density for the minority spin $\rho_{\downarrow}(\omega)$
  for $U=5$, $x=0.5$ and various fields $h$. }
\label{dosdopU5d}
\end{figure}

We now compare the results in the intermediate coupling regime with $U=5$
at half-filling with those at quarter filling, $x=0.5$, where the  Fermi level
falls in the lower Hubbard peak in the spectral density. To see how the
band changes with increasing magnetic field we plot the density of states for
both spin types, for the majority spin electrons  
in figure  \ref{dosdopU5u}  and for the minority spin electrons in 
figure  \ref{dosdopU5d}, for various values of the magnetic field. 

In the majority spin case the lower peak broadens on the low energy side and the
weight in the upper peaks decreases with increase of the field. The opposite
features can be seen in the minority spin case, with the spectral weight in
the lower peak below the Fermi level decreasing and the weight in the upper
peak increasing. Thus the increase of spectral weight below the Fermi level 
for the majority spin electrons, and the decrease for the minority 
spin electrons, can be seen to be due to a
change of band shape rather than a simple relative shift of the two bands,
which would be the case in  mean field theory.
In the fully polarized state there are no minority states
below the Fermi level and the upper peak in the majority state density of
states has disappeared. \par
\begin{figure}[!htbp]
\centering
\includegraphics[width=0.45\textwidth]{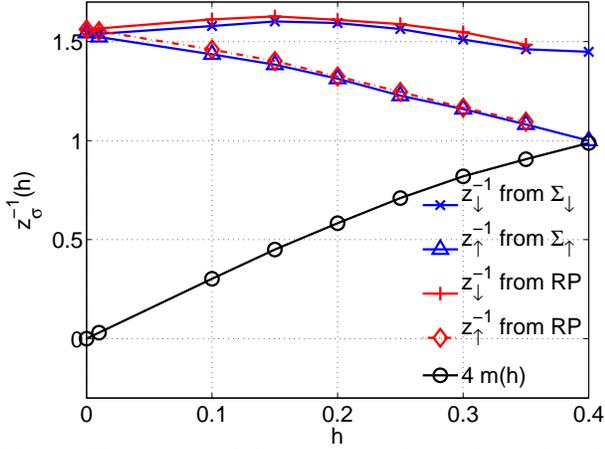}
\vspace*{-0.5cm}
\caption{The inverse of the quasiparticle weight $z_\sigma(h)$ calculated from
  renormalized parameters (RP) and directly from the self-energy for $U=5$, $x=0.5$ and
  various fields $h$. The magnetization $m(h)$ is also displayed.   }
\label{qpweightcompU5x.5}
\end{figure}

The corresponding values for the inverse of the quasiparticle weight
$1/z_\sigma(h)$ are shown in figure \ref{qpweightcompU5x.5} for a range of
fields. As noted in the impurity case, the quasiparticle weights differ for
the two spin types with $z_{\uparrow}(h)>z_{\downarrow}(h)$. The values of
$z_{\sigma}(h)$ have been 
calculated, as described earlier, both from the energy levels (RP) and from 
a numerical derivative NRG derived self-energy. There is reasonable
agreement between the two sets of results, and the small differences to be
seen be attributed to the uncertainty in the numerical derivative of the NRG
self-energy. As in the impurity case, there is an initial decrease of 
 $z_{\downarrow}(h)$ with increase of $h$, whereas
$z_{\uparrow}(h)$ increases monotonically.
 The field
 dependence of the magnetization is also shown in figure
 \ref{qpweightcompU5x.5}, and is similar to the half-filled case with a weak
 interaction ($U=2$).  
We have calculated, but do not show, the corresponding occupation values for
$\tilde n_\sigma^0$ which again agree well with the values of  $\tilde n_\sigma$, verifying
Luttinger's theorem. \par 
Our conclusion from these results, and from calculations with other values of
$U$, is that when there is significant doping, the behavior in the field
corresponds to a weakly correlated Fermi liquid, very similar to that at half-filling
in the weak interaction regime. The only remarkable difference in the field is
the spin dependence of the effective masses as shown in figure
\ref{qpweightcompU5x.5}, which was already found similarly in the impurity case.
\par

\subsection{Near half filling}

\begin{figure}[!htbp]
\centering
\includegraphics[width=0.45\textwidth]{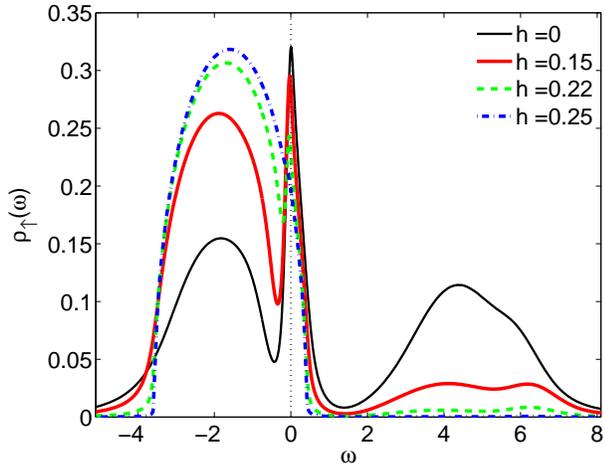}
\vspace*{-0.5cm}
\caption{The local spectral density for the majority spin $\rho_{\uparrow}(\omega)$
  for $U=6$, $x=0.95$ and various fields $h$. }
\label{dosdopU6up}
\end{figure}

\begin{figure}[!htbp]
\centering
\includegraphics[width=0.45\textwidth]{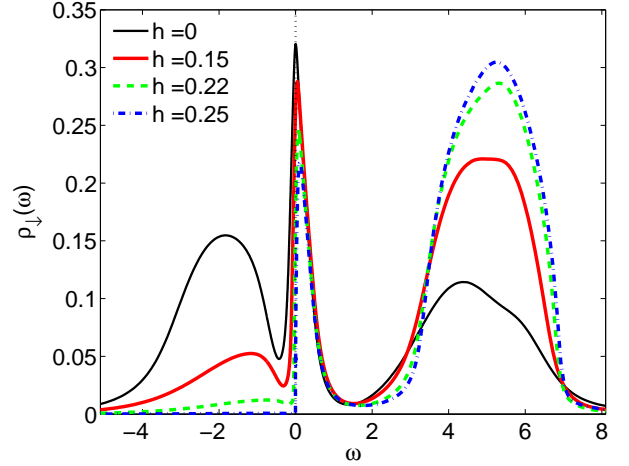}
\vspace*{-0.5cm}
\caption{The local spectral density for the minority spin $\rho_{\downarrow}(\omega)$
  for $U=6$, $x=0.95$ and various fields $h$. }
\label{dosdopU6down}
\end{figure}

Very close to half-filling and for large values of $U$ we have a qualitatively
different parameter regime. Here the system is metallic but we can expect strong correlation
effects when $U$ is of the order or greater than $U_c$, due to the much 
reduced phase space for quasiparticle scattering. We look at the case 
with 5\% hole doping from half-filling and a value $U=6$, which is just
greater than the critical value for the metal-insulator transition. We show
the spectral density of states for both the majority and minority spins states 
and various values of the magnetic field in figures \ref{dosdopU6up} 
 and \ref{dosdopU6down}, respectively. 
There is a clear sharp quasiparticle
peak for $h=0$ at the Fermi level at the top of the lower Hubbard band.
As in the quarter filling case with $U=5$ we see a similar transfer of
spectral weight with increasing field to below the Fermi level for the
majority 
spin case, and above the Fermi level for the minority spins. For large fields
when the system is completely polarized there is still a sharp narrow peak
in the spectral density of the minority spin states above the Fermi level,
though the spectrum for the majority states below the Fermi level is
that of the non-interacting system. 
A spin up electron added above the Fermi
level feels no interaction as the system is completely spin up polarized so these electrons see the
non-interacting density of states. On the other a spin down electron above the
Fermi level interacts strongly with the sea of up spin electrons. The
self-energy due to scattering with particle-hole pairs in the sea creates a
distinct resonance in the down spin density of states just above the Fermi
level. Just such a resonance was predicted by Hertz and Edwards \cite{HE72} for a Hubbard model
in a strong ferromagnetic (fully polarized) state.

\begin{figure}[!htbp]
\centering
\includegraphics[width=0.45\textwidth]{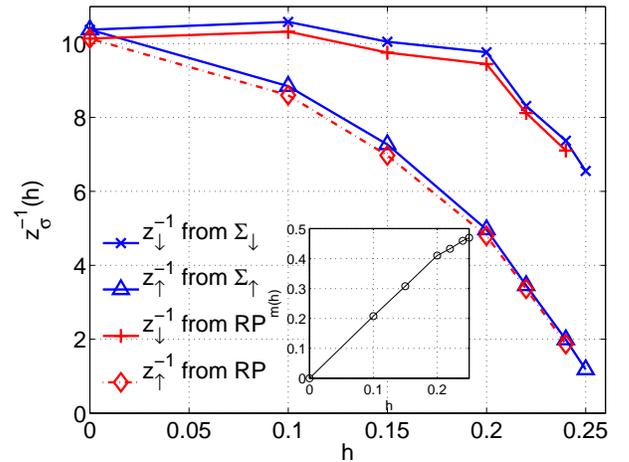}
\vspace*{-0.5cm}
\caption{The inverse of the quasiparticle weight $z_\sigma(h)$ calculated
  from renormalized parameters (RP) and directly from the self-energy for
  $U=6$, $x=0.95$ and various fields $h$. The inset shows the
  magnetization $m(h)$.   }
\label{qpweightcompU6x.5}
\end{figure}

The field dependence of the inverse of the quasiparticle weight is 
presented in figure \ref{qpweightcompU6x.5}. 
Again we find reasonable agreement between the two methods of calculation for these
quantities. The magnetization as a function of $h$ is shown as an inset in the
same figure. The behavior of  $z_\uparrow(h)$ and $z_\downarrow(h)$
as a function of $h$ contrasts sharply with the behavior found for the metallic
state at half-filling with $U=5$ shown in figure \ref{qpweightcompU5}. For
zero field the quasi-particle weight has a very similar value in both cases.
At half-filling the tendency of the magnetic field 
to induce localization resulted in values of $z_\uparrow(h)$ and
$z_\downarrow(h)$ ($z_\uparrow(h)=z_\downarrow(h)$) which decrease sharply
as a function of $h$. In the 5\% doped case with $U=6$, the system 
must remain metallic and the quasiparticles weights, $z_\uparrow(h)$ and
$z_\downarrow(h)$, both increase in large fields though their values differ
significantly. The quasiparticle weight for the minority spin electrons
decreases initially with increase of $h$, whereas that for the majority spins
 $z_\uparrow(h)$ increases monotonically and quite dramatically with $h$.
When the system becomes fully polarized the up spin electrons become
essentially non-interacting, $z_\uparrow(h) = 1$, whereas there is a strong
interaction for a down spin electron and we find in this case
$z_\downarrow(h)\simeq 0.15$. The interpretation for this is as given in the
previous paragraph for the spectral densities.

We conclude that already a small doping of the system is enough to maintain a
metallic character even for very strong interaction. Although the zero field
spectra of the half filled case for $U=5$ and the small doping case with $U=6$
display very similar zero field behavior, i.e. a  strongly renormalized
quasiparticle band with similar $z_{\sigma}$, no field induced localization
transition occurs for finite doping and no metamagnetic behavior is
observed in the latter case. 

\section{Quasiparticle dynamics}
Having deduced the renormalized parameters of the quasiparticles from the NRG
results presented in the previous section, we are now in a position to test
how well we can describe the low energy dynamics of this model in a magnetic
field in terms of a renormalized perturbation theory. We look at the various
parameter regimes in turn. 

\subsection{Free quasiparticle spectral density}
\begin{figure}[!htbp]
\centering
\includegraphics[width=0.45\textwidth]{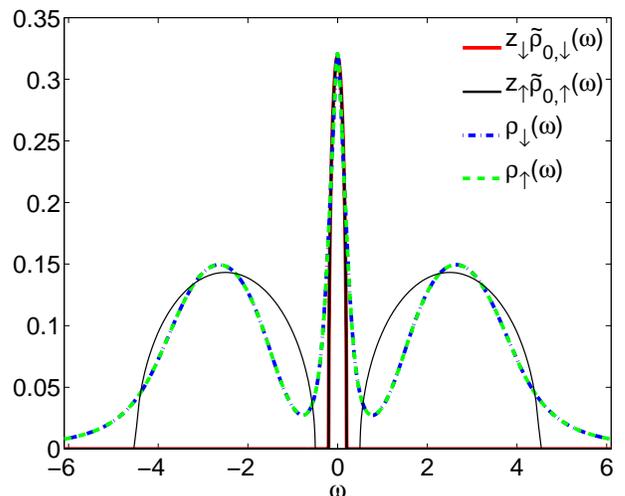}
\vspace*{-0.5cm}
\caption{The free quasiparticle density of states in comparison with interacting local
  spectral density for $U=5$ and $h=0$. We have also plotted a thin black line
  for $\rho_{\rm mf}(\omega)=[D(\omega+U/2)+D(\omega-U/2)]/2$
  which describes the non-magnetic mean field solution and weighted with
  $1-z_{\sigma}$.  }   
\label{qpdosh_0.0}
\end{figure}

\begin{figure}[!htbp]
\centering
\includegraphics[width=0.45\textwidth]{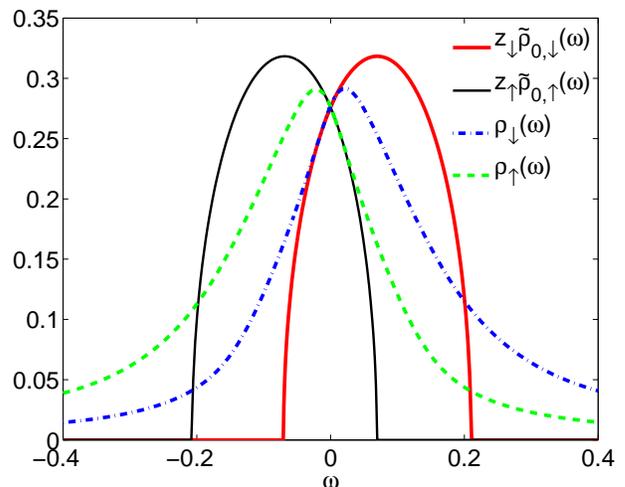}
\vspace*{-0.5cm}
\caption{The free quasiparticle density of states in comparison with interacting the
  local spectral density for $U=5$ and $h=0.15$.  } 
\label{qpdosh_0.15}
\end{figure}

It is of interest first of all to see how  the free quasiparticle density of 
states $\tilde\rho_\sigma(\omega)$ multiplied by $z_\sigma(h)$ compares with
the spectral density $\rho_\sigma(\omega)$. In figure \ref{qpdosh_0.0} we make
a comparison  in the zero magnetic field case. 
We see that the quasiparticle band gives a good representation of the low
energy peak in $\rho_\sigma(\omega)$ and, as expected, does not reproduce the
high energy features. These, however, to a fair approximation can be described
by the mean field solution $\rho_{\rm mf}(\omega)$ weighted with a factor
$1-z_{\sigma}$ as can be seen in figure \ref{qpdosh_0.0}.
A case with a finite magnetic field $h=0.15$, where the peaks in the density
of states of the two spin species are shifted due to the induced polarization 
relative to the Fermi level, is shown in figure \ref{qpdosh_0.15}. 
The figure focuses on the region at the 
Fermi level and one can see the the free quasiparticle density of states
describes well the form of  $\rho_\sigma(\omega)$ in the immediate vicinity
of the Fermi level. It is to be expected that the frequency range
for this agreement can be extended if self-energy corrections are included
in the quasiparticle density of states using the renormalized perturbation theory
as shown in the impurity case \cite{BHO06}.\par
\begin{figure}[!htbp]
\centering
\includegraphics[width=0.45\textwidth]{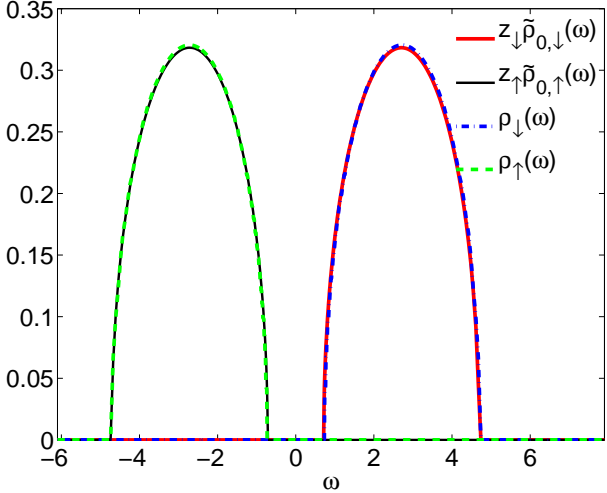}
\vspace*{-0.5cm}
\caption{The free quasiparticle density of states in comparison with interacting local
  spectral density for $U=5$ and $h=0.22$. } 
\label{qpdosh_0.22}
\end{figure}

In the fully polarized case with $h=0.22$ there is complete agreement between the
quasiparticle density of states and  $\rho_\sigma(\omega)$ for both spin types
as can be seen in the results shown in \ref{qpdosh_0.22}, where $z_\uparrow=z_\downarrow=1$.

\begin{figure}[!htbp]
\centering
\includegraphics[width=0.45\textwidth]{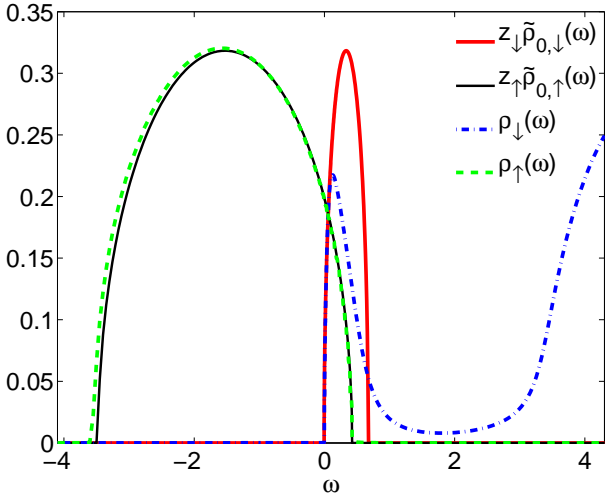}
\vspace*{-0.5cm}
\caption{The free quasiparticle density of states in comparison with interacting local
  spectral density for $U=6$, $x=0.95$ and $h=0.26$. } 
\label{qpdosU6h_0.26}
\end{figure}

In figure \ref{qpdosU6h_0.26} we show a fully polarized result ($h=0.26$) for
the case near half filling, $x=0.95$, $U=6$ discussed in section III.D. 
We can see that the different values for the field dependent quasiparticle weight for up
and down spin $z_{\sigma}(h)$ lead to remarkably different quasiparticle band shapes.
With $z_{\uparrow}\simeq 1$ the majority spin quasiparticle  band is
essentially that of the non-interacting density of states.
The very much smaller value  $z_{\downarrow}$ leads to a narrow quasiparticle
band above the Fermi level. The low energy flank of this quasiparticle band
describes well the narrow peak seen in the spectral density just above the
Fermi level. To describe these strong asymmetries in the spectral densities
near half filling, we need $z_{\uparrow}\gg z_{\downarrow}$, which contrasts
with the cases at half filling such as in figures \ref{qpdosh_0.15} and
\ref{qpdosh_0.22} where always $z_{\uparrow}= z_{\downarrow}$. This suggests
a discontinuous behavior of the renormalisation factors $z_{\sigma}$ on the
approach to half filling.

\subsection{Dynamic susceptibilities at Half-filling}
\begin{figure}[!htbp]
\centering
\includegraphics[width=0.45\textwidth]{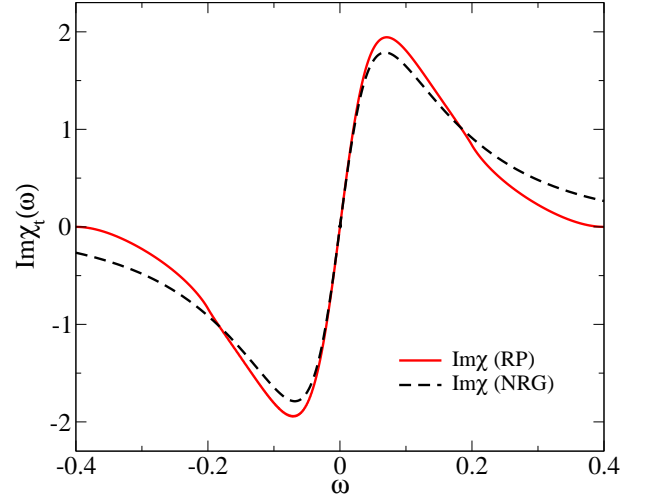}
\caption{A comparison of the imaginary parts of the  transverse dynamic spin
  susceptibility  for $U=5$ and $h=0.0$ calculated using the renormalized
parameters (RP) and from a direct NRG calculation. } 
\label{itchi0.0}
\end{figure}

\begin{figure}[!htbp]
\centering
\includegraphics[width=0.45\textwidth]{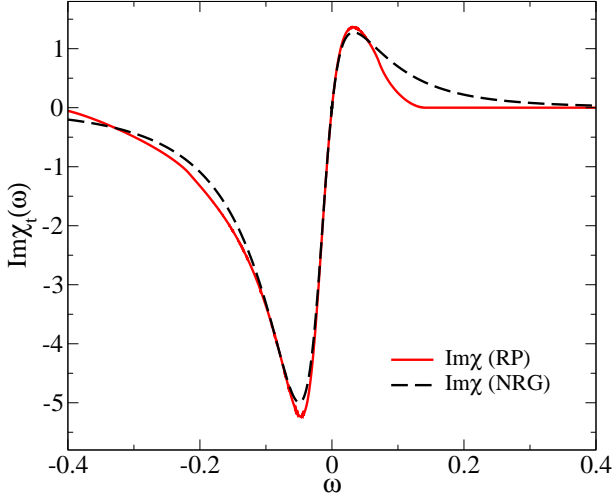}
\caption{Plots of the imaginary part of the  transverse dynamic spin
  susceptibility for $U=5$ and $h=0.15$. } 
\label{itchi0.15}
\end{figure}

We now compare the NRG results for the longitudinal and transverse local
dynamic spin susceptibilities for the same value $U=5$ and a similar range of
magnetic field values with those based on the RPT formulae (\ref{lrrpt})
and (\ref{trrpt}). In figure \ref{itchi0.0} we show the imaginary part of the
transverse spin susceptibility calculated with the two different methods.
It can be seen that RPT formula gives the overall form of the NRG results,
and precisely fits  the gradient of the NRG curve at $\omega=0$.
Some of the relatively small differences between the results might be attributed to the
broadening factor used in the NRG results which gives a slower fall off with
$\omega$ in the higher frequency range, and a slightly reduced peak.
We get similar good agreement between the two sets of results for the same
quantity for the case with a magnetic field $h=0.15$, shown in figure
\ref{itchi0.15}. \par 

\begin{figure}[!htbp]
\centering
\includegraphics[width=0.45\textwidth]{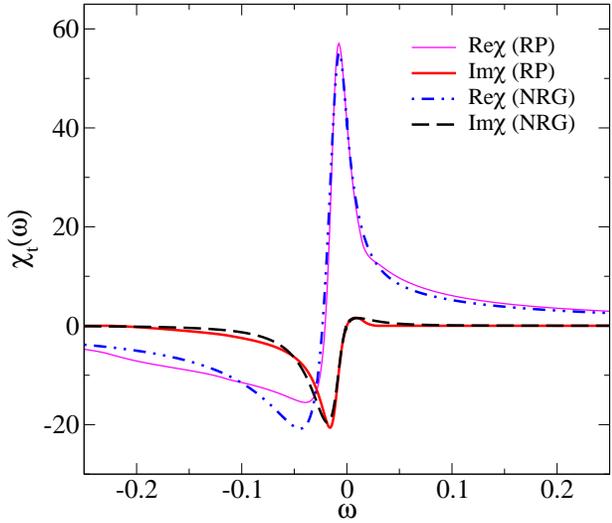}
\caption{The real and imaginary parts of the transverse dynamic spin
  susceptibility  for $U=5$ and $h=0.19$. }  
\label{tchi0.19}
\end{figure}

In figure \ref{tchi0.19}, where we give both the real and imaginary parts of the transverse
susceptibility for $h=0.19$,  we see that this overall agreement is maintained
in the large field regime
where we get metamagnetic behavior. The shapes of the low energy peaks for
both quantities are well reproduced by the RPT formulae. Note that the peak in 
the real part is not at $\omega=0$, so it is not fixed by the condition that
determines $\tilde U_t$, but nevertheless is in good agreement with the NRG
results. Due to their very small values it becomes difficult to calculate 
 $z_\sigma(h)$  as the system approaches localization for larger fields.
In this regime as $z_\sigma(h)\to 0$ the free  quasiparticle density
of states will converge to a delta-function. Self-energy corrections to the
free quasiparticle propagators, which were used in the calculation of
$\tilde\chi_{\sigma,\sigma'}(\omega)$, will become increasing important as this limit is approached.
 Once the system has localized, and is completely
polarized, however, we saw  in figure \ref{qpdosh_0.22}  that the values deduced of $\tilde\mu_\sigma$ 
($z_\sigma(h)=1$) gave the a quasiparticle density of states coinciding
with the NRG result.
\par
\begin{figure}[!htbp]
\centering
\includegraphics[width=0.45\textwidth]{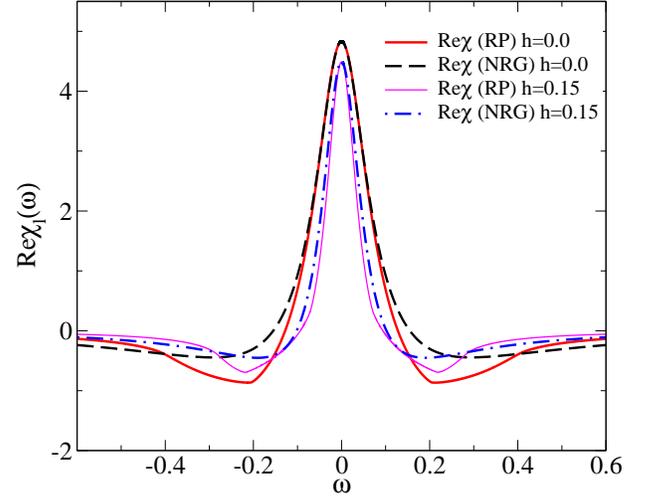}
\caption{The real part of the  longitudinal dynamic spin susceptibility for
  $U=5$ and $h=0$ and $h=0.15$. }  
\label{rlchi}
\end{figure}

\begin{figure}[!htbp]
\centering
\includegraphics[width=0.45\textwidth]{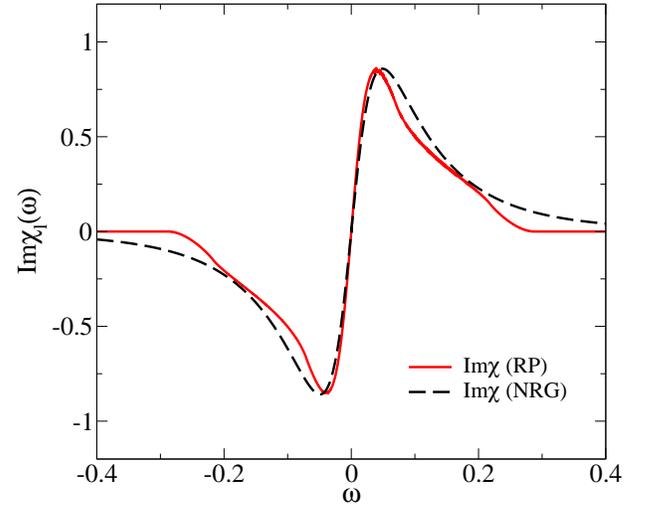}
\caption{The imaginary part of the longitudinal dynamic spin susceptibility
  for $U=5$ and $h=0.15$. }  
\label{ilchi0.15}
\end{figure}

 
Results for the longitudinal susceptibility are shown in figures \ref{rlchi}
and  \ref{ilchi0.15}. In figure \ref{rlchi} we give the values for the real
part as a function of $\omega$ for $h=0$ and $h=0.15$.  
Here the peak height, which is 
at $\omega=0$, is fixed by the condition which determines $\tilde U_l$.
The width of the peaks in the two sets of NRG results, however,  is given
reasonably well by the RPT equations. The imaginary part of the longitudinal
susceptibility obtained by the two methods is given in figure  \ref{ilchi0.15}
for $h=0.15$. Again there is overall agreement between the two sets of results.
The slight undulations seen in the RPT results are due to the sharp cut off
in the band edges in the free quasiparticle density of states.
For larger values of $h$ the agreement with the NRG results is not as good
as that as for the transverse susceptibility, and the central peak in the
real part of the RPT results narrows more rapidly with $h$ than in those 
obtained from the direct NRG calculation.

\subsection{Dynamic susceptibilities at quarter filling}
We give two examples of  results for the susceptibilities for the model at
quarter filling for the case $U=5$ and $h=0.1$. 
\begin{figure}[!htbp]
\centering
\includegraphics[width=0.45\textwidth]{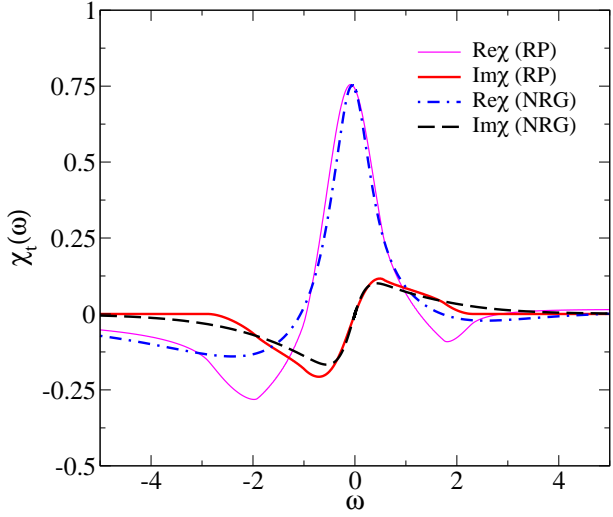}
\caption{The real and imaginary parts of the  transverse dynamic spin
susceptibility  for $U=5$, $x=0.5$  and $h=0.1$. } 
\label{tchix0.5}
\end{figure}
In figure  \ref{tchix0.5}
we give the real and imaginary parts for the transverse susceptibility.
Despite the large value of $U$, we can see that the peak heights are very much
reduced compared with those seen in the half-filled case for $U=5$. The peak
widths are also an order of magnitude larger as can be seen from the 
$\omega$-scale. Nevertheless the RPT results reproduce well the overall features to be
seen in the NRG results.
The real and imaginary parts for the longitudinal susceptibility
are shown in figure  \ref{tchix0.5}. 
\begin{figure}[!htbp]
\centering
\includegraphics[width=0.45\textwidth]{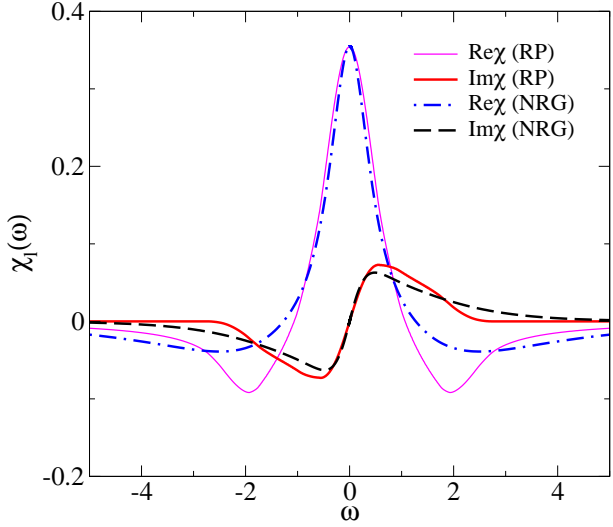}
\caption{The real and imaginary parts of the longitudinal dynamic spin
  susceptibility for $U=5$, $x=0.5$  and $h=0.1$. }  
\label{lchix_0.5}
\end{figure}

Again all the low energy features 
are reproduced in the RPT results. In this regime, apart from the overall
factor of 2, there is less difference between the transverse and longitudinal
susceptibilities than  at half-filling.\par  

\subsection{Dynamic susceptibilities near half filling}

\begin{figure}[!htbp]
\centering
\includegraphics[width=0.45\textwidth]{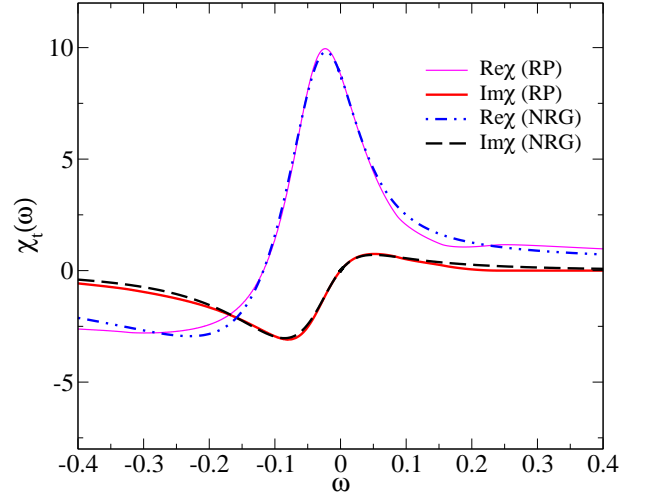}
\caption{The real and imaginary parts of the transverse dynamic spin
  susceptibility  for $U=6$, $x=0.95$  and $h=0.15$. }  
\label{tchix0.95}
\end{figure}

\begin{figure}[!htbp]
\centering
\includegraphics[width=0.45\textwidth]{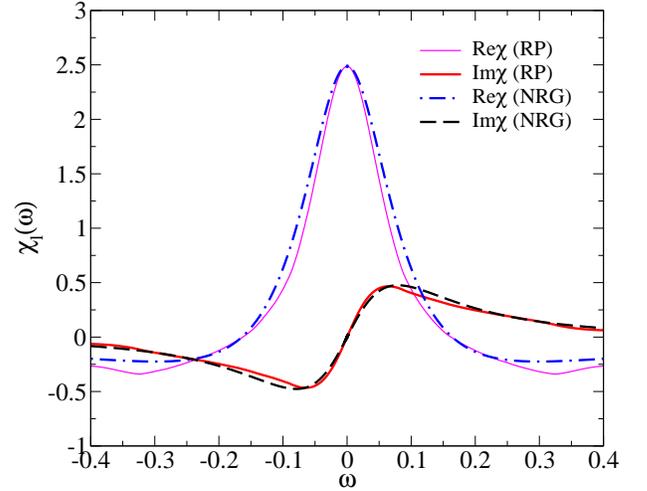}
\caption{The real and imaginary parts of the longitudinal dynamic spin
  susceptibility  for $U=6$, $x=0.95$  and $h=0.15$. }  
\label{lchix0.95}
\end{figure}

Near half-filling ($x=0.95$) we show plots for the two susceptibilities for
parameters $U=6$ and $h=0.15$. In figure \ref{tchix0.95} we give the real and
imaginary parts of the transverse susceptibility. 
The low energy features are
seen on an $\omega$-scale an order of magnitude smaller than that
for quarter filling due to the much stronger renormalization effects in this
regime. There is excellent agreement both with the peak positions and shapes
between the NRG and RPT results for both quantities. This is also seen to be
the case for the real and imaginary part of the longitudinal susceptibility
shown in figure \ref{lchix0.95}, though the peak in the real part can be seen
to be marginally narrower in the RPT results. 

\section{Summary} 
In this paper we have extended our earlier work\cite{HBK06}, where we
described  the low energy 
behavior  of the symmetric  Anderson model in a magnetic field $h$ in terms
of  field-dependent renormalized quasiparticles, to the non-symmetric
Anderson model. The main new feature that emerges is the dependence of
the quasiparticle peak resonance width $\tilde\Delta_\sigma(h)$ 
on the spin type $\sigma$ as well as on the value of the magnetic field.
The $T=0$ spin and charge susceptibilities can be expressed as
explicit formulae in terms of the parameters $\Delta_\sigma(h)$,
$\tilde\epsilon_{{\rm d},\sigma}(h)$ and a local field dependent interaction
between the quasiparticles $\tilde U(h)$. It was also shown earlier
for the symmetric model\cite{Hew06},
that an excellent description of the low energy spin dynamics can be 
obtained by taking into account  repeated quasiparticle scattering
in a renormalized perturbation theory (RPT). We have shown that these
results can also be generalized to the non-symmetric model for the transverse
and longitudinal spin susceptibilities, which again agree remarkably well
with those obtained from a direct NRG calculation.\par

The method for calculating the field dependent quasiparticle parameters
has also been extended here to infinite dimensional lattice models where
the self-energy, as in the impurity case, is a function of frequency only.
We have applied these methods to the Hubbard model
to describe the low energy excitations in terms of a spin dependent
renormalized quasiparticle band. 

For the Hubbard model at half filling, where $z_\uparrow(h)=z_\downarrow(h)$,
we presented results for the three main parameters regimes where the model
displays qualitatively different behavior.
Our results are on the whole consistent with those obtained earlier by  
Laloux et al. \cite{LGK94} obtained using the ED method. An exception is in
the insulating regime for weak fields, where we could not find a convergent
solution of the DMFT equations. We attributed this to the fact that in this regime the
magnetic field is smaller than the exchange coupling between the localized
spins so that ground state would be one in which the spins would have a 
canted antiferromagnetic ordering in the plane perpendicular to the field.

Well away from half filling we find a magnetic response similar to the weakly
correlated case even for large values of $U$. The large phase space for
quasiparticle scattering in this regime leads to modest renormalization
effects. Here, as in the impurity case, we find spin dependent quasiparticle weights,
$z_\uparrow(h)\ne z_\downarrow(h)$. This implies spin dependent as well as
field dependent effective masses, which  have been discussed earlier in work
by Spa{\l}ek et al. \cite{SG90,KSWA95} and Riseborough \cite{Ris06}. 
The calculations by Spa{\l}ek et al. were based on a Gutzwiller \cite{SG90} and a
mean-field slave boson approach \cite{KSWA95}. We can make a 
comparison of our results (section III.D) near half filling, $x=0.95$, with theirs in the
later work \cite{KSWA95}. We find a qualitatively similar behavior with the
majority spin effective mass decreasing with $h$, but quantitatively there are
differences. The field dependence of the minority spin effective
mass $1/z_\downarrow(h)$ shows a
very slow increase initially in both sets of results, but the large field
behavior is quite different. As seen in figure \ref{qpweightcompU6x.5} we
find a significant decrease in $1/z_\downarrow(h)$ for large fields whereas
the corresponding quantity in figure 3 in reference \cite{KSWA95} increases.

The strong magnetic field dependence of the effective masses found in the
calculations by Riseborough is based on the assumption that the system is
close to a ferromagnetic transition (paramagnon theory).
However, DMFT calculations for the Hubbard model find that any ferromagnetism in
the Hubbard model only occurs in a very small region of the parameter space
near half-filling and for very large values of $U$. \cite{ZPB02}
Our results are well away from this regime and the large effective masses
obtained here can
be attributed to the tendency to localization rather than the tendency to
ferromagnetism.

Using the field dependent renormalized parameters $z_\sigma(h)$ and
$\tilde\mu_\sigma(h)$ in the RPT formulae for the dynamic local longitudinal
and transverse spin susceptibilities we found agreement with the overall
features to be seen in the NRG results for these quantities. In the case
of the  transverse spin susceptibility excellent agreement was found
in all the metallic regimes and for all values of the magnetic field
considered, except in the high field regime at half filling  as
the localization point is approached, where consistent values of the
renormalized parameters are difficult to calculate. The comparison of the RPT
results with those from  NRG was also excellent for the longitudinal dynamic
susceptibility in the weaker field regime $h\le 0.15$ but less good for higher
fields, $h>0.15$.\par 

In all metallic parameter regimes a spin dependent Luttinger theorem in the
form $n_{\sigma}=\tilde n^0_{\sigma}$, the number of particles equals the
number of quasiparticles, was found to be satisfied for all strengths of the
magnetic field. In this form it even holds in the fully polarized insulating
state.
\par 

Phenomena like field and spin dependent effective masses and  metamagnetic
behavior have been observed experimentally in several heavy fermion compounds
\cite{AUAO93,GHTYF99,MCCSRC00,DBTWMB06}.
The Hubbard model, however, being a one band model is not an appropriate
starting point to make a quantitative comparison with the heavy fermion class
of materials. 
A periodic Anderson model with a two band structure and including the
degeneracy of the f electrons would be a better model to describe these
materials. Field dependent effects in this model have been studied by several
techniques, modified perturbation theory \cite{MN01}, exact diagonalization
\cite{SI96}, $1/N$ expansion \cite{Ono98} and variational approach \cite{EG97}.
The approach used here could be generalized to the periodic Anderson model,
but restricted to the non-degenerate case and $N=2$ as it is computationally 
too demanding in the NRG to deal with higher degeneracy.
The Hubbard model at half filling has been used as a lattice model to describe
the strongly renormalized Fermi liquid $^3$He \cite{Vol84,LGK94}. However, the
metamagnetic behavior predicted for relevant parameter regime is not seen
experimentally \cite{BFSWRPW98}. 
In section III.D we found for small doping large effective masses, but no metamagnetic
behavior. This raises the possibility that the weakly doped Hubbard model
could serve as a basis for interpreting the experimental results for liquid
$^3$He.

\bigskip
\bigskip
\noindent{\bf Acknowledgement}\par
\bigskip
\noindent
We wish to thank N. Dupuis, D.M. Edwards, W. Koller, D. Meyer and A. Oguri for helpful
discussions and   W. Koller and D. Meyer for their contributions to the
development of the NRG programs. 
One of us (J.B.) thanks the Gottlieb Daimler and Karl Benz Foundation, the
German Academic exchange service (DAAD) and the EPSRC for financial support.

\par
\section{Appendix} 
The free quasiparticle dynamic susceptibility
$\tilde\chi_{\sigma,\sigma'}(\omega)$ for the impurity model in the wide band
limit, $\tilde\Delta_{\uparrow}=\tilde\Delta_{\downarrow}$, were given earlier
\cite{Hew06}. Here we give the more general results for
$\tilde\Delta_{\uparrow}\neq\tilde\Delta_{\downarrow}$,
\begin{equation}
\tilde\chi_{\sigma,\sigma}(\omega)
={-1\over \pi\omega}
{\tilde\Delta_\sigma\over \omega-2i\tilde\Delta_\sigma}\sum_{\alpha=-1,1}{\rm
    ln}\left(1-{\omega\over{\alpha\tilde\epsilon_{d,\sigma}+i\tilde\Delta_\sigma}}\right),
\label{chizz}
\end{equation}
for $\omega>0$, and for $\omega=0$,
\begin{equation}
\tilde\chi_{\sigma,\sigma}(0)
=\tilde\rho_{\sigma}(0).
\end{equation}
The values for $\omega<0$ follow from the fact that
$\Real\tilde\chi_{\sigma,\sigma}(\omega)=\Real\tilde\chi_{\sigma,\sigma}(-\omega)$ and
$\Imag\tilde\chi_{\sigma,\sigma}(\omega)=-\Imag\tilde\chi_{\sigma,\sigma}(-\omega)$.
For $\sigma'\neq\sigma$, 
\begin{eqnarray*}
&&\tilde\chi_{\uparrow,\downarrow}(\omega)= \\
&&\frac{i/2\pi}
{( \omega+\tilde\epsilon_{d,\downarrow}-\tilde\epsilon_{d,\uparrow}
  +i\tilde\Delta_\uparrow-i\tilde\Delta_\downarrow)}
{\rm  ln}\left({\omega-\tilde\epsilon_{d,\uparrow}-i\tilde\Delta_\uparrow\over
 -i\tilde\Delta_\downarrow-\tilde\epsilon_{d,\downarrow}}\right) \\
&&+{i/2\pi\over (\omega+
  \tilde\epsilon_{d,\downarrow}-\tilde\epsilon_{d,\uparrow}-i\tilde\Delta_\uparrow+i
\tilde\Delta_\downarrow)}{\rm
ln}\left({\omega+\tilde\epsilon_{d,\downarrow}-i\tilde\Delta_\downarrow\over   
 -i\tilde\Delta_\uparrow+\tilde\epsilon_{d,\uparrow}}\right) \\
&&+ {-i/2\pi\over (\omega+
  \tilde\epsilon_{d,\downarrow}-\tilde\epsilon_{d,\uparrow}+i\tilde\Delta_\uparrow+i\tilde\Delta_\downarrow)} \times \\  
&&\times \left[{\rm ln}\left(1+{\omega\over
      i\tilde\Delta_\uparrow-\tilde\epsilon_{d,\uparrow}}\right) +{\rm
    ln}\left(1+{\omega\over
      i\tilde\Delta_\downarrow+\tilde\epsilon_{d,\downarrow}}\right)\right] .
\end{eqnarray*}

\bibliography{artikel,biblio1}

\end{document}